\documentclass[3p,times,onecolumn,authoryear]{elsarticle}
\usepackage{amssymb}
\usepackage{subcaption}
\usepackage{amsmath}
\usepackage{amsthm} % definte the proof environment: block letters + italic letters
\usepackage[hidelinks,colorlinks=false]{hyperref}

\usepackage{amsmath,amssymb,amsthm}

\journal{LEM WP}

\makeatletter
\def\ps@pprintTitle{%
 \let\@oddhead\@empty
 \let\@evenhead\@empty
 \def\@oddfoot{\centerline{\thepage}}%
 \let\@evenfoot\@oddfoot}
\makeatother
\begin{document}

\begin{frontmatter}
%\title{Temporal characterization of complexity in agent-based network}
%\title{Extreme events in a minimal model of time-scale free network}
%\title{Anomalous dynamics of a time-scale free network}
\title{A detection analysis for temporal memory patterns at different time-scales}

\author{Fabio Vanni} 
\address{University of Insubria, Varese - Italy}
\ead{fabio.vanni@uninsubria.it}
%\thanks{E-mail: \texttt{fabio.vanni@santannapisa.it}. Corresponding author. }   
 
\author{David Lambert}
% \thanks{E-mail: \texttt{DavidLambert2@my.unt.edu}.} 
\address{Department of Mathematics, University of North Texas, USA} 
\ead{DavidLambert2@my.unt.edu}

%\author{Paolo Grigolini} 
%\address{Center for Nonlinear Science, University of North Texas, USA }
%\ead{Paolo.Grigolini@unt.edu}

%\maketitle
% Abstract; JEL codes, and keywords
\begin{abstract}
	This paper introduces a novel methodology that utilizes latency to unveil time-series dependence patterns. A customized statistical test detects memory dependence in event sequences by analyzing their inter-event time distributions. Synthetic experiments based on the renewal-aging property assess the impact of observer latency on the renewal property. Our test uncovers memory patterns across diverse time scales, emphasizing the event sequence’s probability structure beyond correlations. The time series analysis produces a statistical test and graphical plots which helps to detect dependence patterns among events at different time-scales if any. Furthermore, the test evaluates the renewal assumption through aging experiments, offering valuable applications in time-series analysis within economics.
\end{abstract}

%\begin{abstract}
%The way how  events are observed and revealed influences the statistical property of temporal structure between them. We show how to make use of latency in the data collection procedure to reveal dependence patterns among events at different time-scales.
%In this paper we provide a statistical test for memory dependence in event sequences observing their inter-event time distributions. We then use the renewal-aging property synthetic experiment  to test how much the observer's latency in detecting events  affect the renewal property of the observed distributions.
%%We define and provide a statistical test on the property of symmetrical dependence among a sequence of events generated by some general process.  
%The test uses differences latency in observing the events in order to significantly find evidence of memory patterns in inter-event time intervals at different time-scale of the events. 
%%The statistical evidence make use of the significant  presence of dependency on the probability structure of the events' sequence, and not only in terms of  correlations. 
%We will,finally,  define a novel test on the renewal assumption through the aging experiment on the sequences of events. So we call the procedure the \textit{XA} test and its related plot.
%
%\end{abstract}
%\bigskip
%\noindent  \textbf{Keywords:} Statistical test, renewal processes, bursty events, machine learning technique, \\
%\bigskip
%\noindent \textbf{JEL codes:} C02, C22, C62, G20\\
\end{frontmatter}
\bigskip
\noindent  \textbf{Keywords:} Time-scales memory, Statistical test, Renewal processes, Bursty events, Machine learning technique, Econometrics \\

\section{Introduction}

Latency in counting the events so to have  delayed processes and performing a exchangeability test using time-windows of observations which are period between the initiation of something and the occurrence.
Bursty renewal patterns in evolving systems can be studied  using a temporal-scale perspective of  the inter-arrival event time series, possibly,  revealing  blocks of memory in events. 
Renewal theory as been deeply discussed by the seminal works of \citet{feller1, feller2, cox1967renewal} and it  began with the study of stochastic systems whose evolution through time was interspersed with renewals or regeneration times when, in a statistical sense, the process began anew.
The importance of searching for recurrent patterns is due to the fact that the existence of repetitive scheme makes it always possible to discuss essential features of a sequence of random variables in spite of the laws governing such sequence could be so intricate to preclude a complete analysis. 
For example, the study of recurrent patterns can circumvent the impossibility of a straightforward analysis of  a possible non-markovian behavior of some stochastic processes \citep{smith1958renewal}.

Renewal and regenerative processes are models of stochastic phenomena in which an event (or combination of events) occurs repeatedly over time, and the times between occurrences are independent.
The theory does not need to specify the meaning or effect of  single events, and this is reason why renewal processes are at the core of many stochastic problems found throughout all fields of science.

The field of complex systems can be used as a common framework where many heterogeneous interacting agents produce a systemic bursty dynamics with non-ordinary statistics. In particular,  temporal networks represent a crossroad for many disciplines towards a common understanding of the backbone core of any natural systems. 
%We use the perspective by which  temporal networks can be seen as  static structures with interaction dynamics on links defined as renewal processes with an arbitrary inter-event time distribution but with parameters characteristic to each link. 
In particular the analysis we proposed will be applied to some model of networks to test its validity.

A critical point for processes with renewal patterns is the intrinsic  difficulty to assess if a real world process entails the presence of  such recurrent events over its evolution. consequently, it is of great importance the development of a statistical tool which may detect the presence of renewal events. In literature, this is a challenging issue  has been assessed in different ways. 
A standard statistical tool to detect the presence of renewal events has been determined through a statistical test directly derived by the property of ergodic processes with finite moments of distribution of the inter-arrival times.  The authors \citep{wang2005repairable,bain1991statistical} define different hypothesis tests  to determine whether and how the pattern of events are significantly renewal by analyzing both homogeneous and non-homogeneous Poisson processes.  

Another popular and well known tool is often used in terms of correlation analysis between inter-events time intervals \citep{perkel1967neuronal,perkel1967neuronal2,avila2011nonrenewal}. An important statistics, which quantifies the correlations among events is the serial correlation coefficient (SCC). 

We, in alternatively, propose a statistical test for  renewal processes based on the property of aging of such systems when we observe events a later times of observation. Typical Poisson-types process does not show any aging, on the contrary fat-tails inter-events's distributions show such aging property which typically  makes the previous described tool for renewal assessment useless.

The present aging-based renewal test can also provide deep insights renewal and not-renewal properies of the process for different time-scales so contributing to a better understanding of processes with mixed type of events or the presence of process which behaves differently for different scales, or the presence of truncations in  finite-size systems.

%One  addressed so far through the scaling analysis of the associated diffusion process,  

We will also devote our analysis to the case where a renewal event might be masked by a cloud of secondary events, of Poisson nature, generating the wrong impression that the process is not renewal, and that its memory is a property of the individual trajectories.

In the paragraph 2, we will provide an overview of the importance of the study of renewal process in economics and other sciences, making a review of the key features of renewal process as well as of the aging properties of those.

In paragraph 3, we develop the statistical tool describing the steps needed to test the significant presence of renewal property in the observed processes. We start with synthetic time series whose renewal nature is theoretically known, in order to validate our statistical test.  
In paragraph 4, we apply our test analysis to real world time series  providing variation of the statistical test in the case of data with low number of samples from big data up to single realizations.
%The salient feature of renewal events is that the waiting times between successive occurrences are mutually independent random variables with a common distribution; 

\section{Memory between events and the aging experiment }

Let us consider a counting  process $N(t)$ that counts the number of some type of events occurring during a time interval  $[0,t]$ and   let us suppose $0 \leq t_1 \leq t_2 \leq \ldots $are finite random times at which a certain event occurs.  
The number of the times $t_n$ in the interval $(0, t]$ is:
\begin{equation}
N (t) = \sum_{n=1}^\infty  \mathbf{1}(t_n \leq t),\quad t\geq0
\end{equation}
we will consider $t_n$ as points (or locations) in $\mathbb{R}^+$ with a certain property, and $N(t)$ is the number of points in $[0, t]$. The process $\{ N(t) : t \geq 0\}$, denoted by $N (t)$, is a point process on $\mathbb{R}^+$ . The $t_n$ are its occurrence times (or point locations)\footnote{The point process N (t) is simple if its occurrence times are distinct: $0 < t_1 < t_2 < \cdots$ a.s. (there is at most one occurrence at any instant). }.
The time elapsed  between consecutive events are random variables represent the  inter-occurrence times $\tau_n = t_n - t_{n-1} , \text{ for } n \geq 1$.
The $t_n$ are called renewal times, and $\tau_n$ are the inter-renewal times (or waiting times), and $N(t)$ is the number of renewal events in $[0,t]$.

The epoch of the $n$th occurrence is given by the sum:
\begin{equation}
S_n=\tau_1+\cdots +\tau_n
\end{equation}

As an example. a simple point process $N (t)$ is a renewal process if the inter-occurrence times $\tau_n = t_n - t_{n-1} , \text{ for } n \geq 1$, are independent with a common distribution $\psi$ , where $\psi (0) = 0$ and $t_0 = 0.$  

Those waiting time random variables are called exchangeable if their distribution function is symmetric, so that event series has serial dependence if the value at some time $t$ in the series is statistically equivalent to any other event at another time.

 A finite sequence of  random variables $(\tau_1,\ldots,\tau_n)$ is called exchangeable if $\forall n\geq 2$, 
 \begin{equation}
  (\tau_1, \tau_2, \dotsc, \tau_n)\stackrel{D}{=}(\tau_{\pi (1)}, \tau_{\pi (2)}, \dotsc, \tau_{\pi (n)})  \qquad \forall\pi \in S(n)
 \end{equation}
 where $S(n)$ is the group of permutations of $(1,2,\ldots,n)$ \citep{aldous1985exchangeability,niepert2014exchangeable}. This clearly implies (assuming existence) that means and variances are constant (stationarity). Clearly, independent identically distributed variables are also exchangeable but the opposite is not true in general, for example using the de Finetti's theorem, a exchangeable infinite sequence can be expressed as a mixture of underlying iid sequences \citep{finetti1982exchangeability,shanbhag2001stochastic,kallenberg2005probabilistic}, so  exchangeability is meant to capture symmetry in a problem, symmetry in a sense that does not require independence. Exchangeability generalises the notion of a sequence of random variables being iid and in frequentist approach to statistics obsereved data is assumed to be generated by a series of iid RVs with distribution parameterised by some unknown $p$ which, on the contrary  from a Bayesian perspective, it has some prior distribution, so the random variables which give the data are no longer independent.
 As regard with point processes \citep{huang1990characterization}

Similarly, a time series has serial correlation if the condition holds that some pair of values are correlated rather than the condition of statistical dependence. 
In particular, a sequence of random variables is independent and identically distributed (iid) if each random variable has the same probability distribution as the others and all are mutually independent, i.e. for  $n$ random variables s $(\tau_1,\ldots,\tau_n)$ we have:
\begin{equation}
P(\tau_1, \tau_2,\ldots, \tau_n) = \prod_{i=0}^n P(\tau_i)
\end{equation}
which is the basic property required in the definition of renewal processes.
However, generally,  statistics and in particular machine learning have the purpose of discovering statistical dependencies in data, and the use of those dependencies to perform predictions using the fact that  future observations of a sequence behave like earlier observations. A  formalization of the notion of "the future predictable by past experience" is the exchangeability of random variables.

Finally, we apply our \textit{XA} test to a sequence of events which is not iid but it has the property of exchangeability. 
We replicate a classic process to generate exchangeable binary sequence:  the Polya urn model \citep{hill1987exchangeable}.

\section{Statistical aging experiment}
\begin{figure}[ht]
 \centering
 \includegraphics[width= 0.6\linewidth]{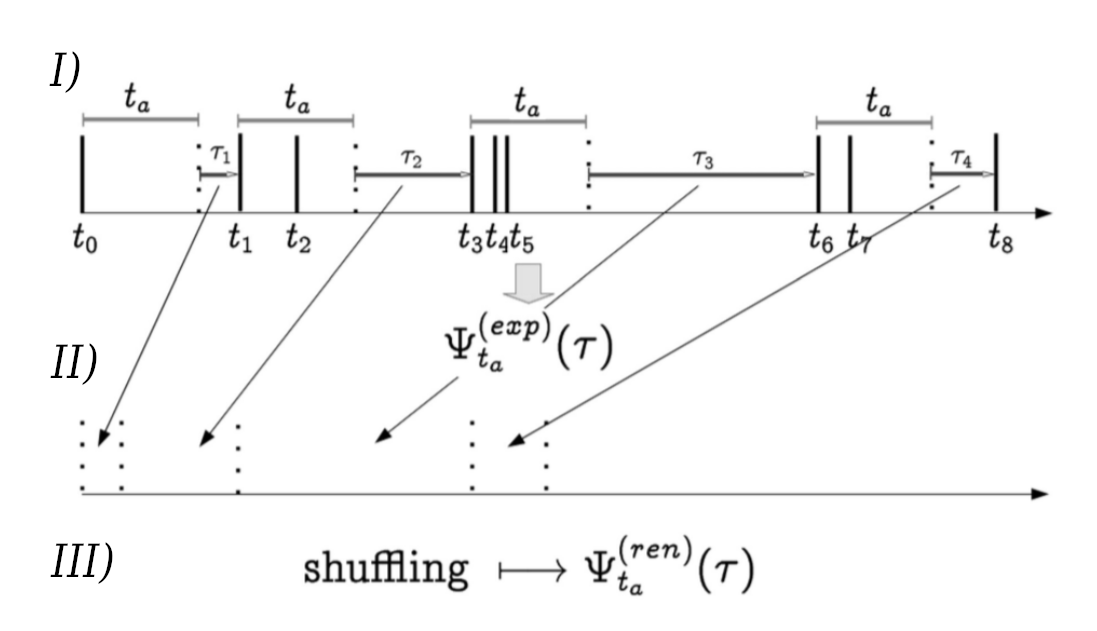} 
 \caption{Illustration of the aging experiment to establish the renewal nature of the process.\textsl{ I)} we age the sequence of events using the observation time $t_a$ from which the observer is ready to detect the next coming event, then the observer cannot detect any other events before another time $t_a$ is passed by, so registering the new aged inter-arrival times $\tau_i$ if the figure. As consequence, the rate of the observation of the process could have impact on the statistical property of the inter-arrival intervals. \textsl{ II)} we collect the new (aged) time intervals which is the aged experimental distribution $\Psi^{(exp)}_{t_a}(\tau)$. \textsl{ III)} is the last step of the aging experiment, we just reshuffle the aged time-intervals so having a distribution $\Psi^{(ren)}_{t_a}(\tau)$ which has to be equal to the experimental distribution if the process is renewal.}
 \label{sketch}
  \end{figure}

To turn the theoretical prediction that would make it possible to establish renewal aging though an ensemble observation, we have to find a way to establish renewal aging observing a single sequence. Fig. \ref{sketch} illustrates how to make the renewal aging assessment using a single realization. We move a window of size $t_a$ along the time series, locating the left size of the window on the time of occurrence of an event. The window size prevents us from assessing if there are or not events before the end of the window. We record the time distance between the end of the window and the occurrence time of the first event that we can perceive. 
 The moving window serves the important purpose 
 of mimicking the use of a very large number of identical systems. In fact, if non-stationarity is not due to changing with time rules, the exact moment when an event occurs can be selected as time origin of the observation process. Beginning  our observation process at a distance $t_a$ from the occurrence of an event can be done with the events of the time series under study. This is the purpose of the moving window of Fig. (\ref{sketch}).

 Using the intermittence jargon we call \emph{laminar region} the time interval between the occurrence of two consecutive events.  It is evident that the times that we record are portions of the original laminar regions.

 In this case the aging experiment illustrated by Fig. (\ref{sketch}), generating only fractions of the original laminar region, has the effect favoring the long-time laminar regions, because cutting a very large laminar region may have the effect of leaving very extended also the laminar region produced by the delayed observation. The short-time laminar regions are affected much more from the delayed observation. 

\section{Renewal-Aging  test}
In order to asses a statistical measure of the renewal patterns in a sequence of events, we biuld our hypothesis upon a well define aging experiment in the previous paragraph.

We finally consider the problem of multiple testing of a single hypothesis, with a standard goal of combining a number of p-values without making any assumptions about their dependence structure. We will use a combined probability test to combine the results from several independent tests bearing upon the same overall hypothesis. 

\begin{enumerate}
\item $\forall t_a$ latency, a point-wise significance test analysis using the aged distributions and the reshuffled version:
\begin{itemize}
\item[a)] perform the two/sample test  techinque (i.e. Kolmogorov-Smirnov or Permutation test) to verify the hypothesis that the the original aged sequence and a shuffled aged one have the same distribution (null hypothesis)
\item[b)] Check if the distribution of the $p$-values  obtained by the test are uniformly distributed  and compute Fisher's combined $p$-value as age-wise significance test. 
\end{itemize}
\item Perform a $p$-value boxplot over different ages $t_a$ (latency) for a qualitative overview of the renewal property for each $t_a$. We also computes  the geometric means of $p$-values for every age $t_a$ 
\item As a global statistical evidence one can test the property of the behavior of geometric-mean for each box  respect to the expected distribution under the hypothesis the process is renewal.
\end{enumerate}

First we discuss the test for a given age $t_a$ for which an aging experiment has been carried out. 

\subsection{Two samples tests}

We want to statistic quantifies a distance between the empirical distribution functions of the two kind of aged-samples derived from the aging procedure previously explained: the original aged sequence of inter-arrival times and the reshuffled aged sequence one.

One of the central goals of data analysis is to measure and model the statistical dependence among random variables.
Empirical distribution functions have been used for studying the serial independence of random variables at least since Hoeffding [62].

The discussion about two-sample tests also applies to the problem of testing whether two random variables are independent. The reason is that testing for independence really amounts to tasting whether two distribution are the same, namely, the joint distribution and the product distribution.

There are many statistical tools which can provide such hypothesis testing, and we will focus the attention to the well know Kolmogorov-Smirnov test (K-S) which is a non-parametric distribution free statistical hypothesis  procedure for determining if two samples of data are from the same distribution \citep{Kolmogorov33,Smirnov33}\footnote{There are many alternatives which also improve the K-S test, for example the Anderson-Darling test or  The Cramer-von Mises test, but we use the K-S as main reference of our Renewal hypothesis as a standard and well assessed procedure in the statistical literature.}

The combination of many K-S test is performed through a such prescription can be developed for independent  observations under the same hypothesis, so it can be applied for artificial data from model simulations and from big data analysis where many  independent measurements has been performed over the same process, as different neurons during their spiking activities. 

Let us indicate the aged time interval sample as $m$ i.i.d. random variables  $(\tau_{t_a}^{(1)},\ldots ,\tau_{t_a}^{(m)})$ and let us indicate  another independent sequence obtained where the aged time interval have been shuffled  as $n$ i.i.d. random variables  $(s_{t_a}^{(1)},\ldots ,s_{t_a}^{(n)})$. Let  $\mathcal{T}_m(z)$  and   $\mathcal{S}_n(z)$ the corresponding empirical distribution functions and the new random variable $D_{m,n}$ by  
\begin{equation}\label{eq_teststat}
D_{m,n}=\sup_z| \mathcal{T}_m(z) - \mathcal{S}_n(z) |
\end{equation}
and using the  Glivenko-Cantelli theorem \citep{van2000asymptotic,gibbons2011nonparametric} which guarantees that  the two empirical distributions have samples made up from the same distribution,  the  statistic $D_{m,n}$ almost surely converges to zero.   Such test statistic is   appropriate for a general two sided hypothesis test:
\begin{align}
\mathcal{H}_0 \, :\; & \psi_{\tau}(z) =  \psi_s(z) \qquad \text{ for all } z \qquad  (\textit{exchangeable events}) \\
\mathcal{H}_1\, : \; & \psi_{\tau}(z) \neq  \psi_s(z) \qquad \text{ for some } z \quad  (\textit{not-exchangeable events})
\end{align} 

The p-value for  statistic $D_{m,n}$ may be obtained by evaluating the asymptotic limiting distribution $Q(t)$ as:
\begin{equation}\label{eq_pvalue}
p=\text{Pr}(D_o \leq D_{m,n}| \mathcal{H}_0)  
\end{equation}
where $D_o$ is the observed value of the two-sample K-S test statistic, as consequence we obtain the value that is the probability that the observed statistic occurred by chance alone, assuming that the null hypothesis is true. 

In order to get the value of eq.\eqref{eq_pvalue} one could use the analytical approach used in \citep{feller2015kolmogorov} where:
\begin{equation}
\lim_{m,n \to \infty}\text{Pr}\left(D_{m,n}\sqrt{\frac{n+m}{nm}} \leq z \right ) = 1-2\sum_{i=1}^{\infty}(-1)^{i-1}e^{-2i^2z^2}=: Q(z)
\end{equation}
where $Q(z)$ is the c.d.f of Kolmogorov-Smirnov distribution and so $D_{m,n}$  serves as a  consistent test statistic for our  hypothesis test.

Following the approach in \citet{stephens1970use}, we can  numerically determine the $p$-value as $p \simeq 1-Q(\lambda)$  where $\lambda = D_{o}\cdot \left(\sqrt{(n+m)/nm}  +0.12  +0.11/(\sqrt{(n+m)/nm} \right)$ which  becomes asymptotically accurate as $nm/(n+m) \geq 4$   \footnote{This is a numerical procedure used in most of K-S algorithm in the main scientific programming language based on the \citet[ch.14]{press2007numerical} in C, Matlab, STATA, R and many others. An alternative approach is by comparing the test statistic $D_{m,n}$ with a critical value $c_{\alpha}$ where $\text{Pr}(D_{m,n}\geq c_{\alpha} | \mathcal{H}_0) \leq \alpha$, so obtaining the rejection decision if $D_o>c_{\alpha}\sqrt{(n+m)/nm}$ where the critical values $c_{\alpha}$ can be obtained from tables.}.

Let us notice that we will also make use of computational approaches to testing   statistical   hypotheses such as two sample permutation test especially useful when the assumptions of K-S test are violated:  for example K-S test is exact only  for continuous variables. but it is conservative for discrete variables, so in the case of small samples the non-continuous variables have a significant effect on the test, in alternative we will make use of computational statistical tests as the permutation test approach. 

Another violation of K-S family tests is the case when the two samples are not mutually independent or when the sample are not completely random. In those cases we will devote a discussion in how to detect and try to avoid or minimize this sort of artifact dependence among data.

The K-S test is originally used to asses if a single observation can fit with the hypothesis of a renewal sequence in a single realization. However, in the case we have many independent sequences we can perform multiple  hypothesis testing for the renewal assumption. 
This is the case  when the sequences are synthetic realizations derived from models so it is always possible to perform as many tests we want so to have a more reliable outcome of the $p$ values about the renewal property of the underlying process. Another situation in which we can perform multiple testings is in the presence of big amount of data made up of independent observations of the same (or at least equivalent) process where. for example,  we can ran a statistical test on each gene in an organism, or on demographics within each of hundreds of counties keeping the tests independent among them.

  In those cases  the challenge would be to find a suitable procedure to combine the results from several independent tests bearing upon the same overall hypothesis (renewal assumption)\footnote{Such research question would be distinguished from another type of multiple hypothesis testing of  statistical comparison of many competing hypotheses in order to discover hidden processes underlying observed patterns of data (called data dredging or p-hacking). }.

\subsection{Meta-analysis}
Once we have chosen the statistical test to assess the equivalence between the two distributions derived from the aged-sequence and a reshuffled one, one can produce many two-sample comparisons producing many $p$-values obtained through a chosen two-sample test (K-S in this case). 
Combining p-values from independent statistical tests is a popular approach to meta-analysis, in particular we will introduce a  procedure of combining the information in the $p$-values from different renewal statistical tests in order to obtain a single overall test under the assumption that the tests are statistically independent. There are many methods  for combining $p$-values in  a single test of common hypothesis as extensively shown by \citet{loughin2004systematic}.

Basically, our analysis is based on  the approach in  Fisher \citep{fisher1932statistical} about a consistent way to combine $p$-values coming from independent repeated tests over the same null hypothesis.

Consider a set of $N$ independent hypothesis tests, each of these to test a certain null hypothesis $\mathcal{H}_{0|i}, i=\{1, 2, \ldots, N\}$. For each test, a significance level $ p_{i}$ (p-value) is obtained. All these $p$-values can be combined into a joint test whether there is a global effect, i.e., if a global null hypothesis $\mathcal{H}_0$ can be rejected. The test is based on the fact that the probability of rejecting the global null hypothesis is related to intersection of the probabilities of each individual test. If the underlying test statistics $D_1 , \ldots , D_N$  have absolutely continuous probability distributions under their corresponding null hypotheses, the joint null hypothesis for the $p$-values is $\mathcal{H}_0 : p_i \sim U[0,1]$: so the several $p$-values are considered as random variable which is uniformly distributed when the global null hypothesis is true.

  \begin{figure}[!ht]
   	 \centering
   	  	                  \includegraphics[width=0.65\linewidth]{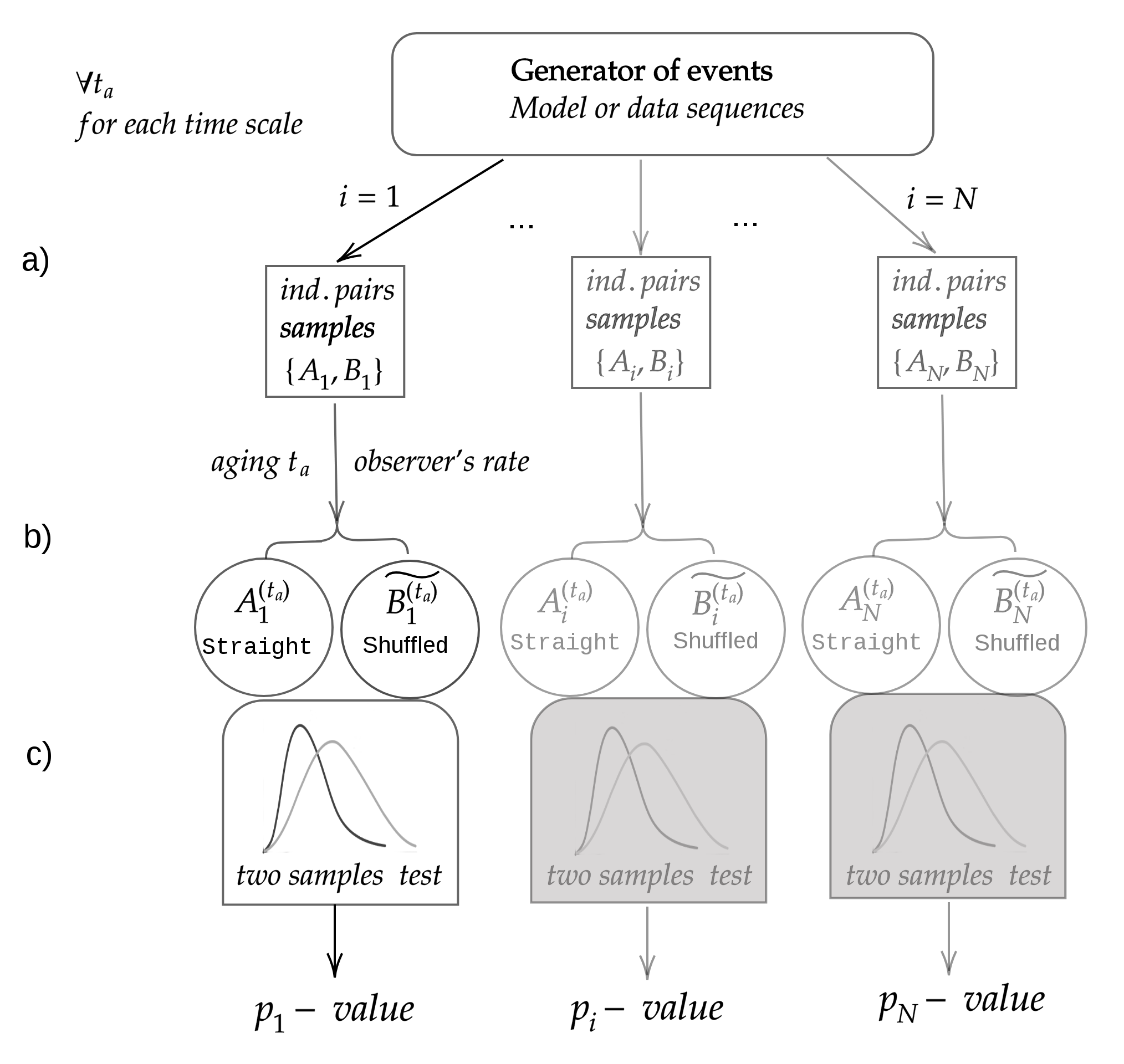}
   	                        \caption{Meta-analysis of \textit{XA} test per age. Per each observation rate $t_a$, in step \textsl{a)} we can generate $N$ pairs of realizations of the process (synthetic ones or observations from data), $A_i,B_i$ which are independent. In step \textsl{b)} the aging experiment of the two inter-arrival times is performed, in one case (i.e. the sequence). In step \textsl{c)} A two sample statistical significance is performed (e.g. K-S test) and so computing the $p$value of the \textit{i}-th comparison test. After collecting a vector of $p$-values $\{p_1,\ldots, p_N \}$ which have to be uniform distributed under the null hypothesis there is no memory in the system (the process is renewal). The procedure will be repeated for different time scales $t_a$ so aging the observed distributions}
   	                        \label{fig_scheme} 
   \end{figure}

Let us to stress here that the geometric mean of a set of $p$-values is $g_p= \left( \prod_{i=1}^{N}p_i \right) ^{1/N}$, no matter how alike or different between the individual elements, so the geometric mean  is not technically a combined  $p$-value but it is the "best" average for the $p$-value \citep{vovk2018combining}.

We will use the Fisher's approach for a qualitative and quantitative statistical clarity of our renewal hypothesis testing, having in mind that each test is performed for different ages $t_a$, so that the overall test is spread over all the possible $t_a$, where a pure renewal processes would always accept the null renewal hypothesis for any $t_a$.

%\begin{figure}[!ht]
%      \centering
%     	 \begin{subfigure}[c]{0.45\textwidth}
%     \includegraphics[width=1\linewidth]{img/powerlaw_R1.png}
%       	                        \caption{{ R=0.1 } }\label{}
%           \end{subfigure}%      
%              %\hspace{2\textwidth}
%      	       %\vspace{20pt}
%      	       \qquad
% \begin{subfigure}[l]{0.45\textwidth}
%      	 \centering
%      	                  \includegraphics[width=1\linewidth]{img/powerlaw_R2.png}
%      	                  \caption{{  R=0.9   }}
%      	                  \label{}
%      \end{subfigure}
%      \caption{ { Complementary cumulative distribution for the inter-event time between two consecutive cycles.  Higher minimal policy requirement keeps the boom-bust cycles more confined to an epxonential rate in the long period. In fig (a) we observe the clear extended power law , meanwhile in (b) we have a strong truncation for poissonian decay due to finite size effect. In the inset we can see the clear exponential nature of the cutoff.}}
%       \label{fig_cutoffR}
%\end{figure}

In order to consider an combined measure of many independent $p_i$ p values, a test based on the geometric mean is  a preferable since it is consistent, in the sense that  it can not fail to reject the overall test null hypothesis although the result of one of the partial tests is extremely significant.

 Under the null hypothesis of renewal assumption, let us call $N$ the number of $p_i$-values from $N$ independent K-S tests, under the null,  the geometric mean of uniformly distributed $p$-values has a probability density function as
 
 \begin{equation}\label{eq_pdf_geo}
 \rho_N(g_p) = \frac{N}{\Gamma(N)} (-N g_p \log g_p)^{N-1} \mathbb{I}_{(0,1)} (g_p)
 \end{equation}
 
So, under the null hypothesis,  it is expected that the geometric mean variable has the following mean and variance:
\begin{align}
E[g_p]&=\mu_0=(1+1/N)^{-N}  \\
& \to e^{-1} \quad  \text{ for } N\to \infty \nonumber \\ 
\text{Var}[g_p]& = \sigma_N^2=(1+2/N)^{-N} - (1+1/N)^{-2N}  \\
&\sim e^{-2}/N + O(1/N^2) \to 0 \quad  \text{ for } N\to \infty  \nonumber
\end{align}

\subsection{Overall time-scales test :   \textit{XA} plots}
In the previous paragraphs we first defined the tools to compare two aged distributions coming from ordinary and shuffled inter-arrival time intervals. We performed many repeated independent tests obtaining many $p$ values for each repetition of the K-S test which we combine in unique average $p$ value for each time-scale (age) $t_a$.

Finally, in the last step, we perform the same repeated K-S tests for different length of observation time $t_a$ viewing at the geometric mean $g_p(t_a)$ for different temporal-scales  of agings.

\begin{figure}[!ht]
   	 \centering
   	  	                  \includegraphics[width=0.65\linewidth]{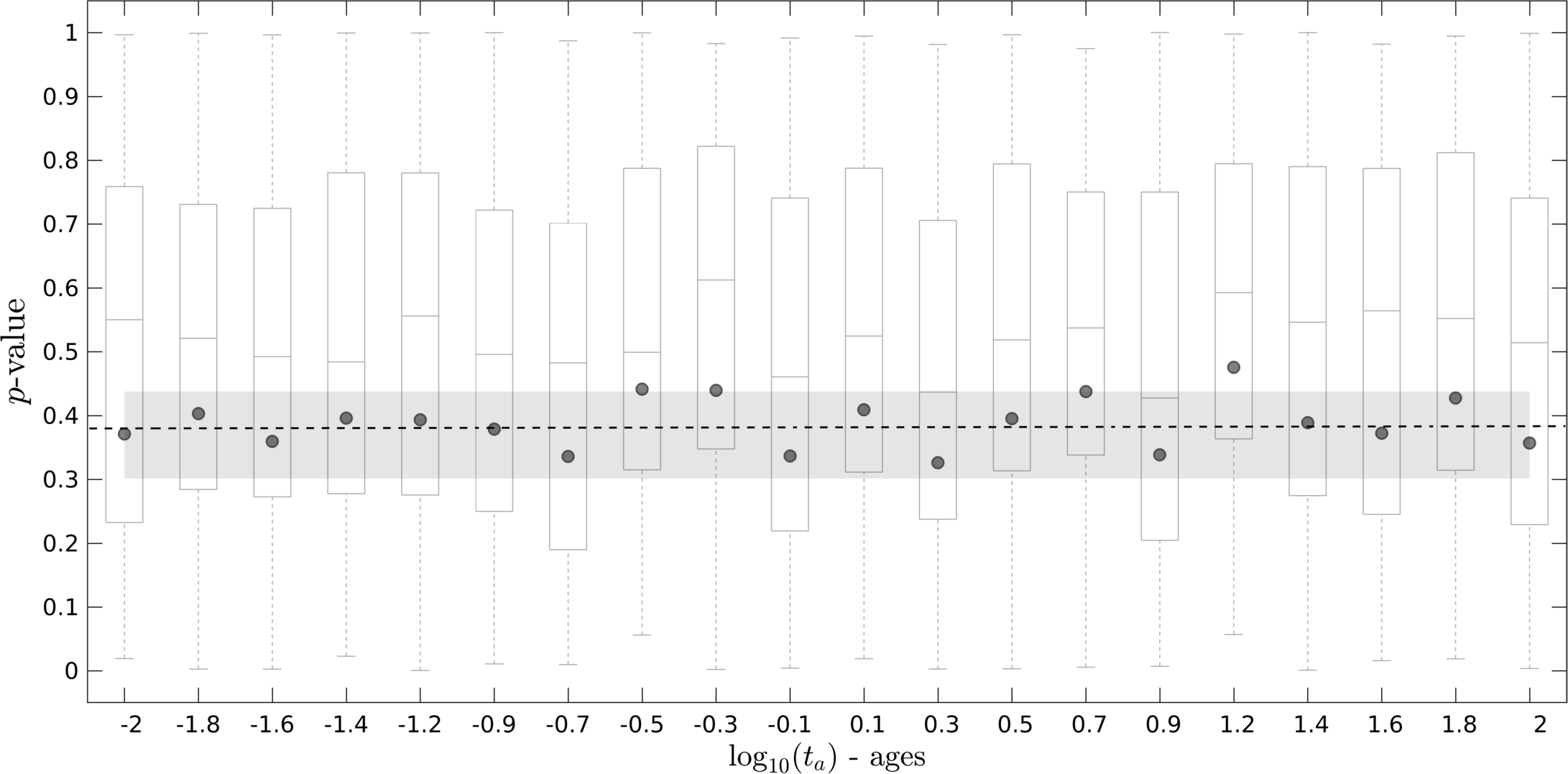}
   	                        \caption{R-A plot for a renewal exponential inter-arrival times. Since the process is renewal each time-scale $t_a$ is the age at which the repeated p-values are evaluated in the $N=100$ Kolmogorov-Smirnov tests. One can see the boxplot for each $t_a$ which is a uniform distribution so that the geometric mean is a random variable  which stays within the $95\%$confidence interval in the gray stripe around its mean value $1/e$, the gray horizonatla dotted line. As results the test cannot reject the null hypothesis of renewal events so not revealing significant presence of memory between events at all the time-scales of the process.}
   	                        \label{fig_pure} 
   \end{figure}

At this purpose we can construct the Renewal-Aging (R-A) plots which shows  the geometric mean points  of p-values over different ages, then a stripe is shown which indicates a $95\%$ confidence interval around the expected geometric mean. If the computed $g_p(t_a)$'s are statistically compatible with the renewal assumption those points stays within the stripe with the correspondent  expected geometric mean. Moreover light gray bars are shown for each $t_a$, those are the blox plots showing the distribution of p-values over $N$ different K-S test for each particular $t_a$. In Fig.\ref{fig_pure} we show the typical R-A plot for 20 ages $t_a$. Under the null hypothesis one epxect to see a uniform distributied box-plot where for each age $t_a$ one sees a uniform distribution of p-values for serveral K-S tests, and the geometric mean should be around its expected value within a certain expected deviation.

A process is intended to be "pure" renewal or "pure" not-renewal if  it exhibits always the same inter-events dynamics for every time scales $t_a$, as in the case of exponential and power-law distribution presented here. In such cases it is possible to derive a final unique index of significance of the global renewal test for every ages $t_a$.

Despite the R-A test is quite simple it is important to discuss the meaning of some parameters in the test as reported in Table \ref{tab_par} . First of all we can fix a smallest and largest ages which gives the minimal scale of memory and the maximal one in the sequence of events. In particular the minimal $t_a^{min}$ is fixed so that we have observation rates which can acctualy can aged the observed sequence of events. The maximum $t_a$ instead can be considered as proportional to length of the sequence multiplied by the  average rate of the waiting times between two consecutive events. This is done in order to have enough samples to perform the two-sample comparison test at large ages. 
The temporal resolution $T_a$ is the number of ages $t_a$ and it is connected to the resolution of temporal scales of the \textit{XA} plots since it is connencted with the increments between two consecutive ages. There is no limit to such parameters since it always increase the number of geometric means for which we make the \textit{XA} test.  
Finally the number of K-S test for each $t_a$ is kept constant for all the ages and it sets the precision of the \textit{XA} test since increasing $N$ we have that, under the null,  the standard error of geometric means tends to zero, so that  the  amount  of  chance  fluctuation  we can expect in sample estimates will reduce. This number $N$ represents also the degree of freedom in the Fisher's combined test so we decided to keep this name, and its only constraint for such parameter is the computational speed of the test.

\begin{table}[!ht]
\centering
\begin{tabular}{|c|c|c|}
\hline
   $t_a^{max}$      & $T_a$&  $N$\\ \hline \hline
  $\sim L / \langle \tau \rangle$ & (\textit{free}) &  \textit{dof}   (\textit{free}) \\ \hline
        largest    &    \textit{temporal resolution} &  \textit{statistical precision} \\ % \hline
     memory block      &  test's sample size  & population sample \\ \hline
\end{tabular}
\caption{Parameters of the \textit{R-A} test. Where the length $L$ is the number of events in the data sequence $\{ \tau_{i} \}_{i=1,\ldots , L}$. Then,  $T_a$ is the number of $t_a$'s used in the repeated tests so that $\{ t_{a_{j}} \}_{j=1,\ldots, T_a}$. Moreover, $T_a$ also represents the number of geometric means calculated in the whole statistical procedure, and $T_a$  is considered the sample size of the test  We can also notice that another parameter striclty related to $T_a$ is the step-size $\delta _{t_a} =t_a^{max}/T_a $ which represents the sampling interval between two consecutive geometric means in the \textit{R-A} plots. On the other side, $S$ is related to the variability of the test's sample, as consequence increasing $N$ we reduce the variance of the geometric mean variables so increasing the statistical precision of the test. } \label{tab_par}
\end{table}

A global test statistics  is the standard core of the sample mean of  $\{ g_p(t_a)  \} $:
\begin{equation}
Z_g=\frac{\overline{g_p} - \mu_0}{\sigma_n  }=\frac{\overline{g_p} - 1/e}{\sqrt{e^{-2}/N}  }= \sqrt{N}\, (e\,\overline{g_p}-1)
\end{equation}
and $Z_g \to 0$ for $n\to \infty$ under the null hypothesis, otherwise, under the alternative hypothesis, $Z_g$ diverges. 

For large samples, the test statistic is approximately distributed as a standard normal distribution according to the central limit theorem. Therefore, using a lower-tailed test we can reject the null hypothesis of a renewal process if $Z_g<z_{-\alpha}$ where $\alpha$ is a given confidence level  accepting the alternative hypothesis of memorry between the events,  whenever $Z_g\leq \mu_0 +z_{1-\alpha}\sigma_n$. Otherwise, a upper-tailed test given $Z_g>z_{\alpha}$  can detect positive dependence in the samples we used in the test which has nothing to do with a possible correlation in the sequence of events. Such positive dependence,for example can arise if one uses not independent samples in the K-S tests or a poor performance of K-S procedure if the sample size is low (discrete samples when the continuous sample approximation fails). However this case indicates an artifact in the test which has to be taken in account and the presence of such spurious artifact should be discussed in details separately.

It is important to evaluate the effect of the test parameter $N$ (number of trials in repeated two-sample statistical tests)  and $T_a$ (number of time scales $t_a$ ages), since the have an impact on the power of the test in accepting the presence of memory in the event sequence when the renewal hypothesis is false. The power of a lower-tailed z-test is:
 \begin{align*}
 \text{Power}
 	& = Pr(Z_g\le \mu_0 + z_{1-\alpha}\,\sigma_n|H_1) \\
 	& = 1-Pr\left(\frac{\displaystyle Z_g-\mu_1}{\displaystyle\sigma_n}\ge \frac{\displaystyle \mu_0-\mu_1}{\displaystyle\sigma_n}+ z_{1-\alpha} \Big\vert H_1\right)\\
 	& = 1-\Phi\left(\frac{\displaystyle \mu_0-\mu_1}{\displaystyle\sigma_n}+ z_{1-\alpha}\right)
 \end{align*}
 where $ \Phi$ is the cdf of a normal distribution where we used $z_{\alpha} = - z_{1-\alpha}$, and $H_1$ is the alternative hypothesis of non-renewal process with the consequent presence of memory between the events.
 
 It is worth to point out how  $N$  (p-value trials on repeated two-sample tests) and $T_a$ (the sample size of the geometric mean i.e. number of time-scales $t_a$) are two free parameters which can be chosen in order to get a desired spatial and temporal  resolution respectively. If they are increased also the power of the test increases since it increase the ability to reject the null hypothesis when the null hypothesis is false, so revealing the presence of memory when the assumption of lack of memory is actually false.
 
\begin{figure}[!ht]
       \centering
      	 \begin{subfigure}[c]{0.45\textwidth}
      \includegraphics[width=1\linewidth]{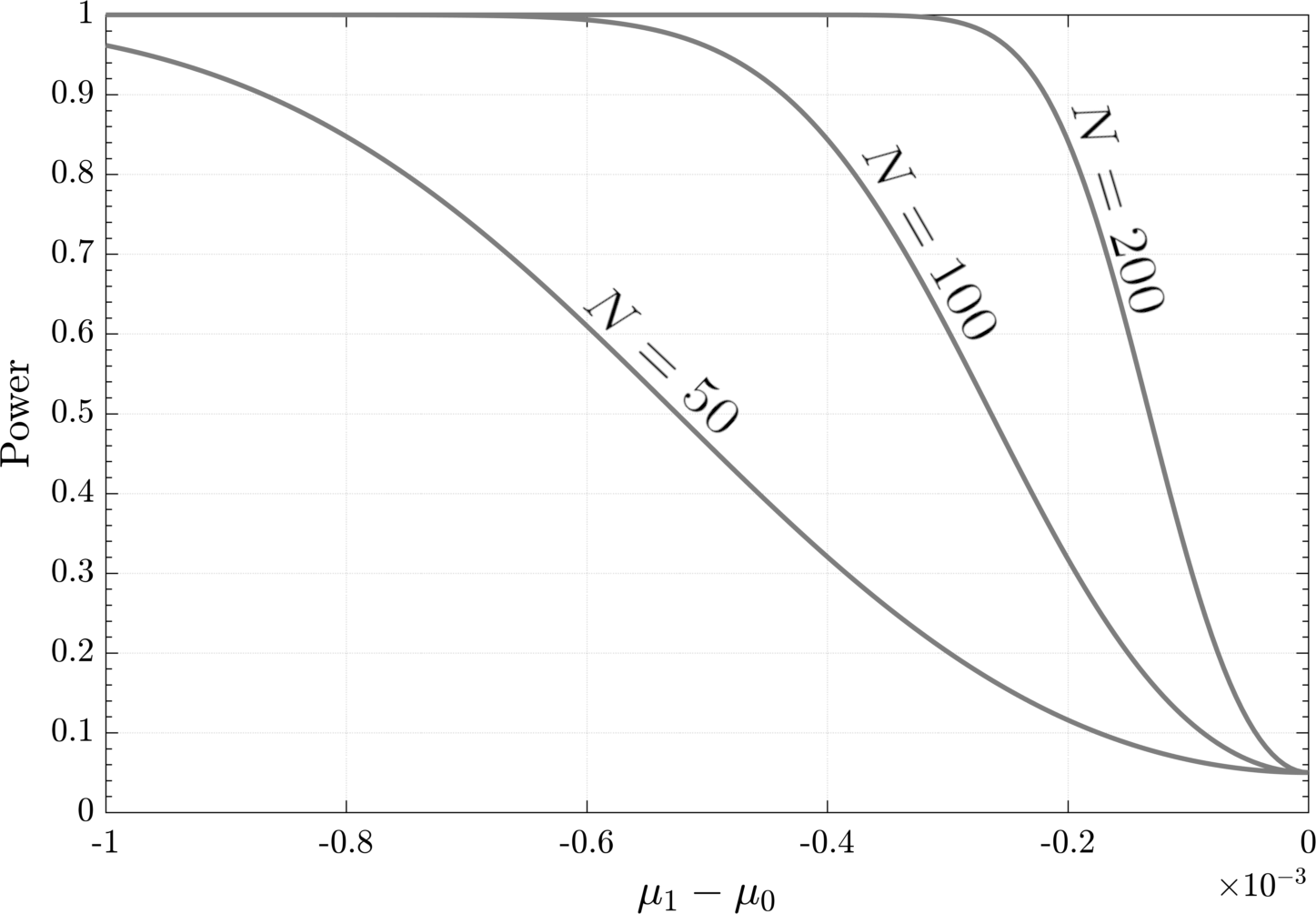}
        	                        \caption{ Power of the test respect to various values of $N$ for a fixed $T_a=100$. }\label{}
            \end{subfigure}%      
               %\hspace{2\textwidth}
       	       %\vspace{20pt}
       	       \qquad
  \begin{subfigure}[l]{0.45\textwidth}
       	 \centering
       	                  \includegraphics[width=1\linewidth]{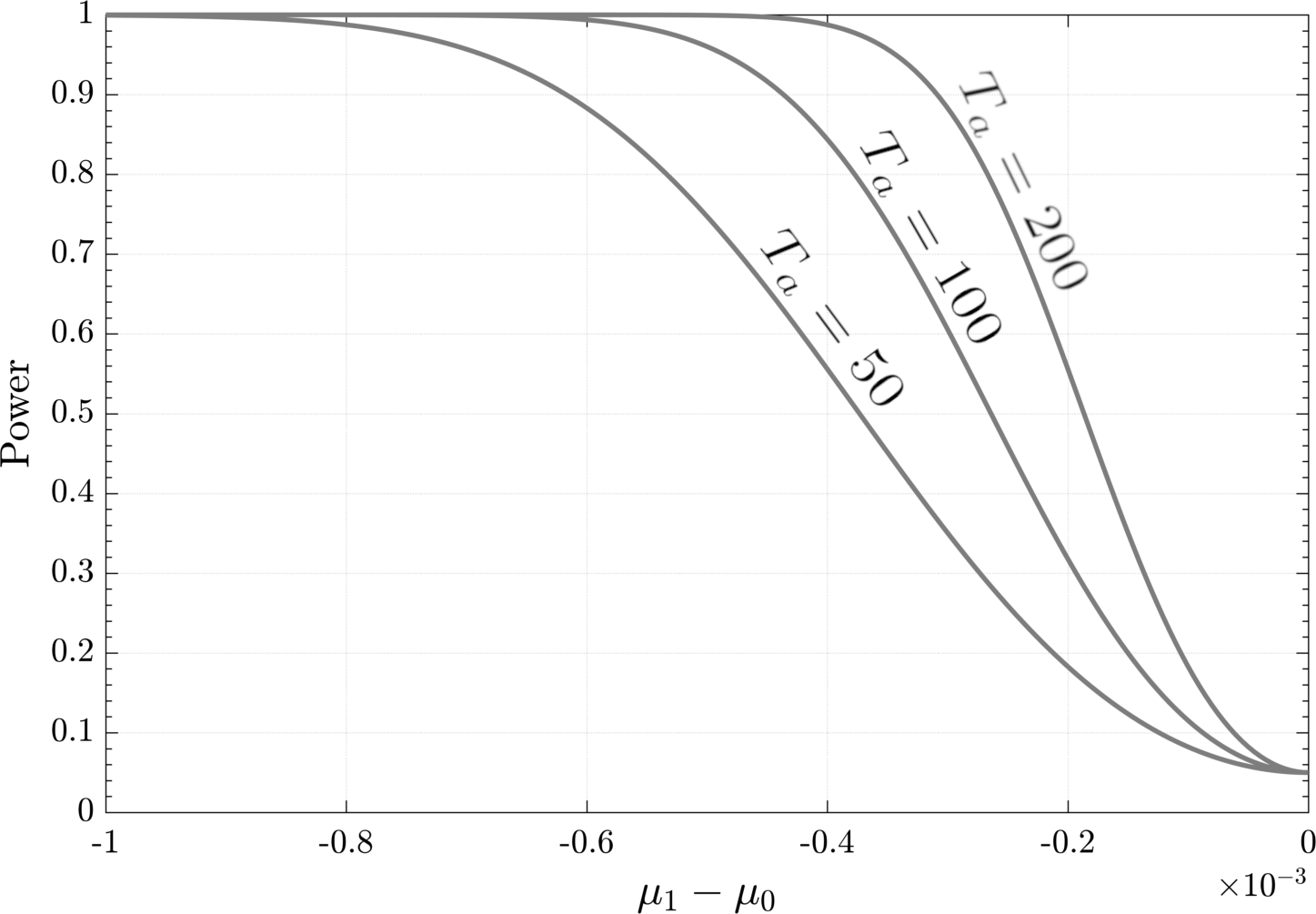}
       	                  \caption{ Power of the test respect to various values of $T_a$ for a fixed $N=100$.}
       	                  \label{}
       \end{subfigure}
       \caption{ Power of lower one-tailed Z test of XA plots at $5\%$ confidence level.The probability of rejecting the renewal hypothesis given that the alternative non-renewal hypothesis is true since memory between events is significantly present. In the figure (a) we plot the power of the z-test for different number of total number of repetition of K-S test for single time-scale $t_a$ with a total number of $T_a=100$ ages. In fig (b) we plot, instead, the power of the test in the case we increase the number of time-scales $T_a$ for a given $N=100$.}
        \label{fig_power}
 \end{figure}

 In Fig.\ref{fig_power} we plot the power of the z-test for different values of the parameters $N$ and $T_a$ revealing that both increases the power of the test, so in principle one should prefer to use large values for those parameters in order to detect the presence of memory.

 However, the \textit{XA} plots can reveal also an heterogeneous behavior of the process at different time-scales of the observer's rate, so a straight Z-test is not suggest without first considering the \textit{XA} plot in its complete version, before proceeding with a  final test on the overall hypothesis.
 
Another useful test for assessing the overall renewal hypothesis  is the Maximum Likelihood estimation of our sample of geometric means fitted respect to a normal distribution.
So we can set a change of variable to transform the geoemtric mean $\rho_N (x)$ distribution into a normal distribution $G(y)$ with a given mean $\mu_y$ and variance $\sigma_y ^2 $. We can find a transformation $y=h(x)$ with the Jacobian factor so that the  transformation is $G(y)=\rho_N(x)\frac{1}{|\rho'_N (x)| }$. 

\begin{figure}[!ht]
       \centering
      	 \begin{subfigure}[c]{0.425\textwidth}
      \includegraphics[width=1\linewidth]{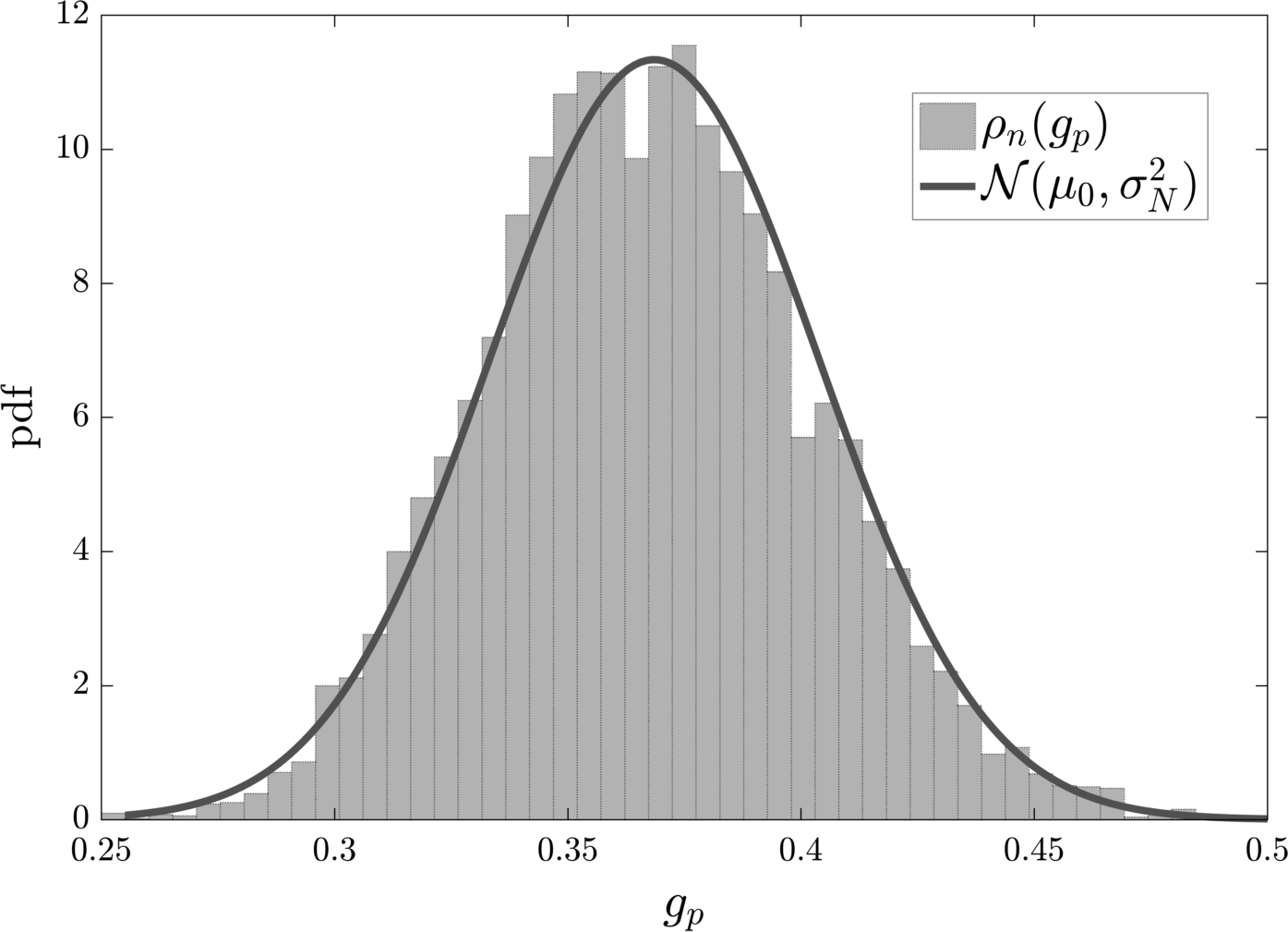}
        	                        \caption{Geometric mean distribution and its transformed gaussian variable with same mean and variance.}\label{}
            \end{subfigure}%      
               %\hspace{2\textwidth}
       	       %\vspace{20pt}
       	       \qquad
  \begin{subfigure}[l]{0.425\textwidth}
       	 \centering
       	                  \includegraphics[width=1\linewidth]{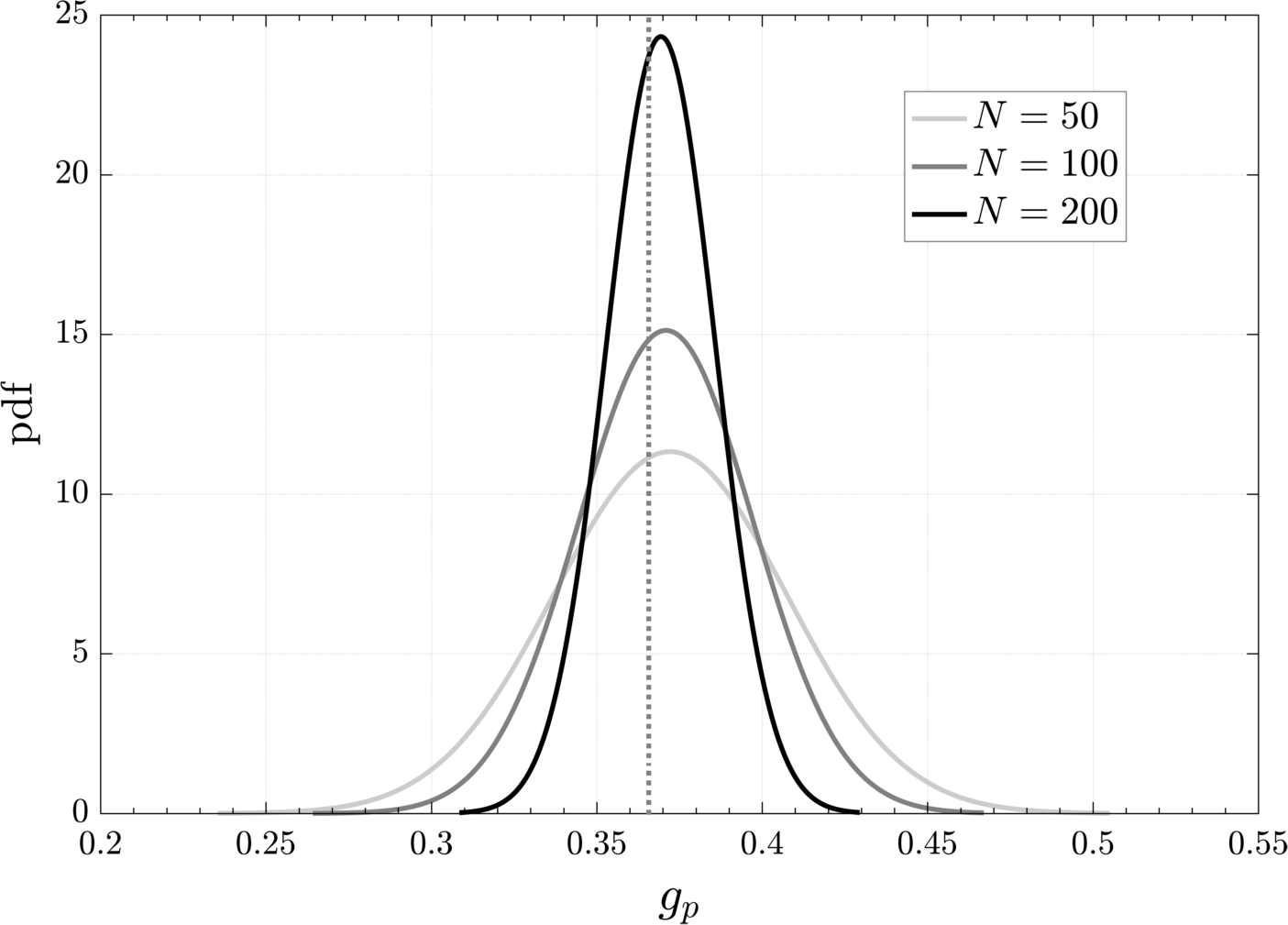}
       	                  \caption{The transformed gaussian variable for different values of two-sample tests. }
       	                  \label{}
       \end{subfigure}
       \caption{Distribution of the geometric mean variable $g_p$ and its transformed gaussian variable under the function $h(\cdot)$ for different values of repeated trials $N$. As shown in (b), for $N\to \infty$ the expected value is $e^{-1}$ (the dotted vertical line) and zero variance. }
        \label{fig_geomean}
 \end{figure}

The final solution of such transformation is:
\begin{equation}
h(x)=\sqrt{2\pi \sigma_y ^2} \text{ erf}^{-1}\left[ \pm \frac{2\Gamma(N, -N\log(x))}{\Gamma(N)}  \right] + \mu_y
\end{equation}
Using the new Gaussian random variables of the geometric means we can perform a fit to the expected theoretical  normal distribution, see Fig.\ref{fig_geomean} where the transformation is computed in order to have a normal distribution with the same mean and variance of the original geometric  mean distribution.

However whatever  one wants to use, z-test or MLE-normal fit,  it is always strongly suggested  to not use those tests  by itself but it has to be always associated to the graphical inspection of the \textit{XA} plots in order to detect memory in events at different time scales.

\clearpage
\section{Validation of the \textit{XA} test}
In this section we will validate the R-A test on synthetic realizations of events' sequences derived from models from which we know to be renewal or not-renewal by the mathematical property of the process. 
We wil apply the statistical technique to renewal, not-renewal and mixed processes, to highlight the effectiveness and usefulness of R-A plots as general tools to detect memory  between events.

We, so, provide a couple of example where the sequence of inter-event times are dependent but not correlated.

First, let us consider a simple, auto-correlated  volatility structure which  can generate a sequence of samples which mimics dependent time-intervals  which are not-correlated.

Let $z_t$ be a i.i.d. mormally distributed random  variables, $z_t\sim \mathcal{N}(0,1)$, and let $\sigma _t$ follow an $AR(1)$ process  $\sigma_t = \beta \sigma_{t-1} + s\,\epsilon _t$ where $ \epsilon  \sim \mathcal{N}(0,1)$.  Finally, let us define the time-intervals as:
\begin{equation}\label{eq_depend}
\Delta_t=\exp\{\,z_t\, \sigma _t\}
\end{equation}
where $\Delta_t$ and $\Delta_{t-1}$ are uncorrelated but clearly dependent.

\begin{figure}[!ht]
	\centering
	\begin{subfigure}[c]{0.475\textwidth}
		\includegraphics[width=1\linewidth]{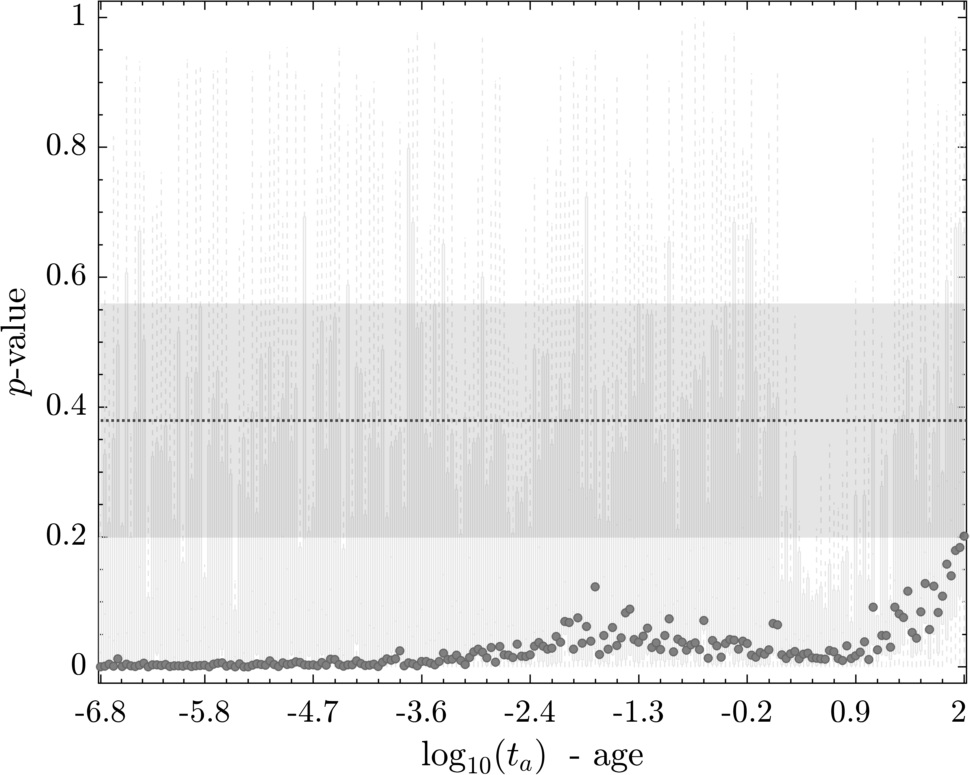}
		\caption{\textit{XA}-plots. The test clearly reveals a significance presence of memory in the inter-event times which clearly shows dependency.}\label{}
	\end{subfigure}%      
	%\hspace{2\textwidth}
	%\vspace{20pt}
	\qquad
	\begin{subfigure}[l]{0.475\textwidth}
		\centering
		\includegraphics[width=1\linewidth]{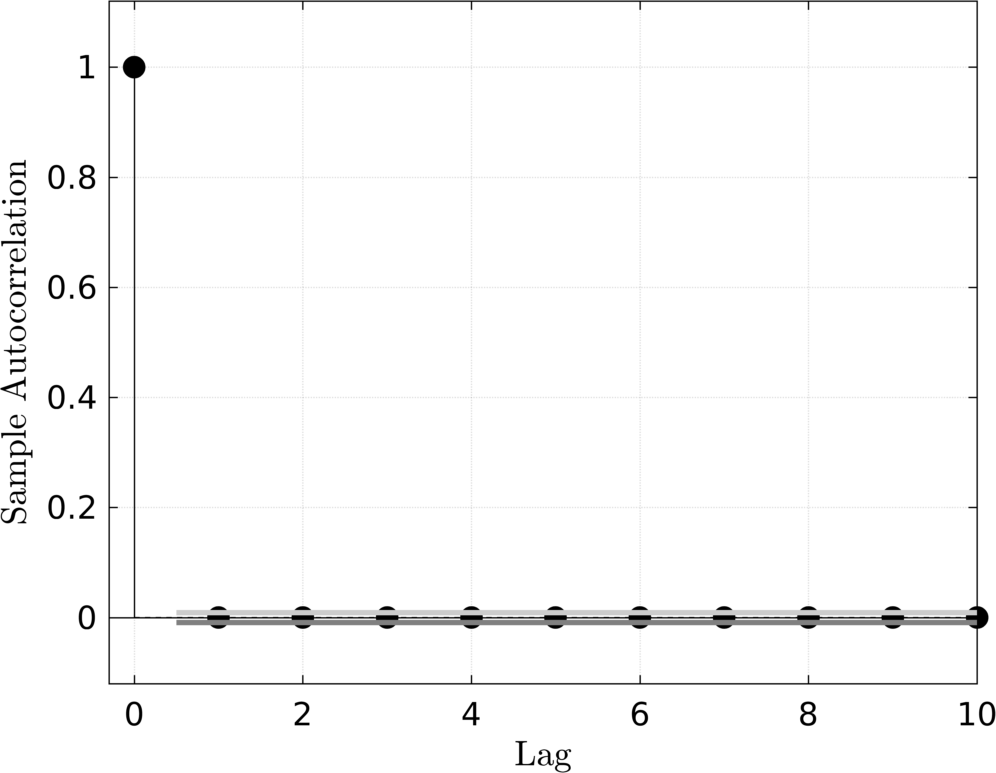}
		\caption{ACF. The sample autocorrelation function does not reveal any significant autocorrelation patterns in the sequence of inter-event times within the confidence bound}
		\label{}
	\end{subfigure}
	\caption{Statistical test for memory of inter-events time intervals from eq.\eqref{eq_depend} with $b=0.97, s=0.89$ in terms of dependency (a) and in terms of correlation only (b) }
	\label{fig_XApool}
\end{figure}

If series values are independent, then nonlinear instantaneous transformations such as logarithms, exponential, absolute values, or squaring preserve independence.

However, the same is not true of correlation, as correlation is only a measure of linear dependence. For example, in financial time series analysis, higher-order serial dependence structure in data can be explored by studying the autocorrelation structure of the absolute returns (of lesser sampling variability with less mathematical tractability) or that of the squared returns (of greater sampling variability but with more manageability in terms of statistical theory). If the returns are independently and identically distributed, then so are their transformations. Hence, if the absolute or squared returns admit some significant autocorrelations, then these autocorrelations provide evidence against the hypothesis that the returns are independently and identically distributed.

\subsection{Synthetic Renewal Processeses}
\begin{figure}[!ht]
   	 \centering
   	  	                  \includegraphics[width=0.6\linewidth]{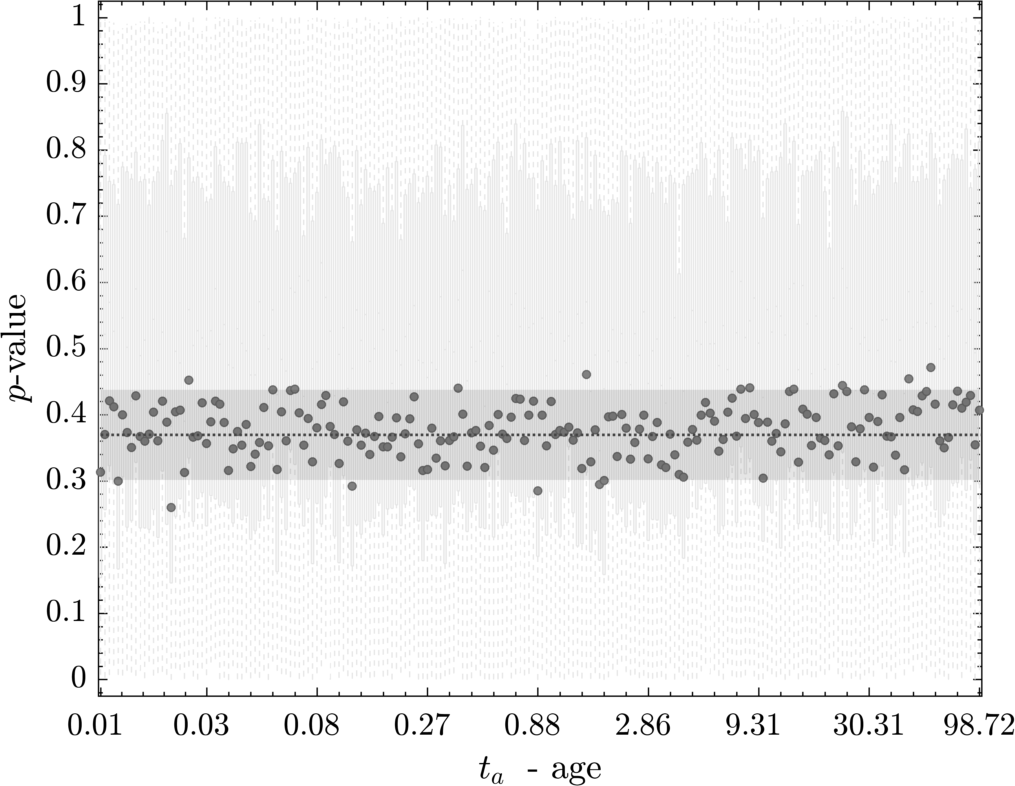}
   	                        \caption{Renewal hypotheis testing via aging experiment for a poissonian point process with exponential inter-arrival time intervals with rate $1$. we can notice that the renewal property is never rejected}
   	                        \label{fig_expren} 
   \end{figure}

As preliminary example, we use an homogeneous Poisson process whose inter-arrival times are exponential distributed and the events are renewal. 
The inter-arrival times has distribution $\psi (\tau) = \lambda e^{-\lambda \tau}$ and  the aged distribution $\psi_{t_a}(\tau) = \lambda e^{\lambda t_a} e^{-\lambda \tau}$, since it is a renewal point process we expected to not reject the null hypothesis for all the $t_a$ ages.  In Fig. \ref{fig_expren} we plot the aging renewal hypothesis testing on a process with characteristic time $\lambda =1$  performed over $N=100$ K-S independent tests. 
The boxplot represents the distribution of the $p_i$ values which are uniformly distributed  for every $t_a$. 

   In practice, the statistic requires a relatively large number of data points  to properly reject the null hypothesis \citep[ch.6]{conover1999practical}. As a consequence, in setting the domain of $t_a$ one have to take in account the length of the observed time intervals sequence $T$ in such a way tht the number samples of the two distributions in the K-S tests should have ${nm}/(n+m) >4$ and $\min \{n,m\}>30$  in order to make the K-S test to work properly. In our case we have a total number of events for the samples is $ n  \approx \lambda T =3\cdot 10^3 $,  so the maximum age is $t_a^{max}= n/30 = 100$

\begin{figure}[!ht]
       \centering
      	 \begin{subfigure}[c]{0.475\textwidth}
      \includegraphics[width=1\linewidth]{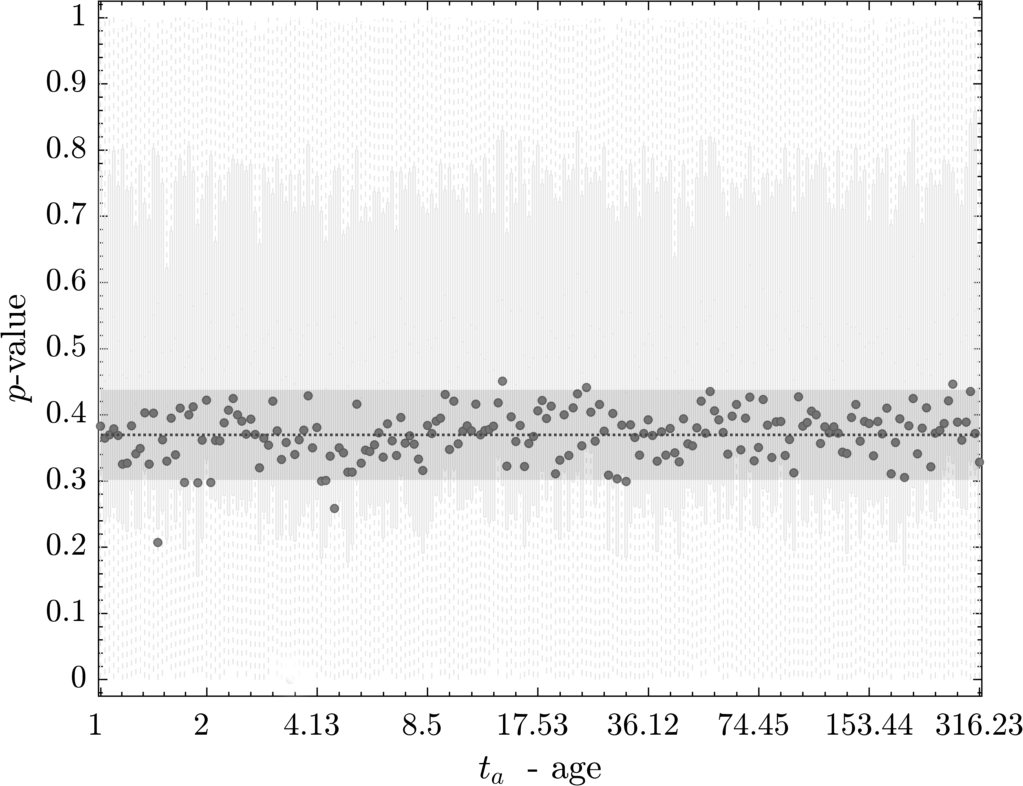}
        	                        \caption{ Power Law distributed waiting times for events with $\mu=2.1$. }\label{}
            \end{subfigure}%      
               %\hspace{2\textwidth}
       	       %\vspace{20pt}
       	       \qquad
  \begin{subfigure}[l]{0.475\textwidth}
       	 \centering
       	                  \includegraphics[width=1\linewidth]{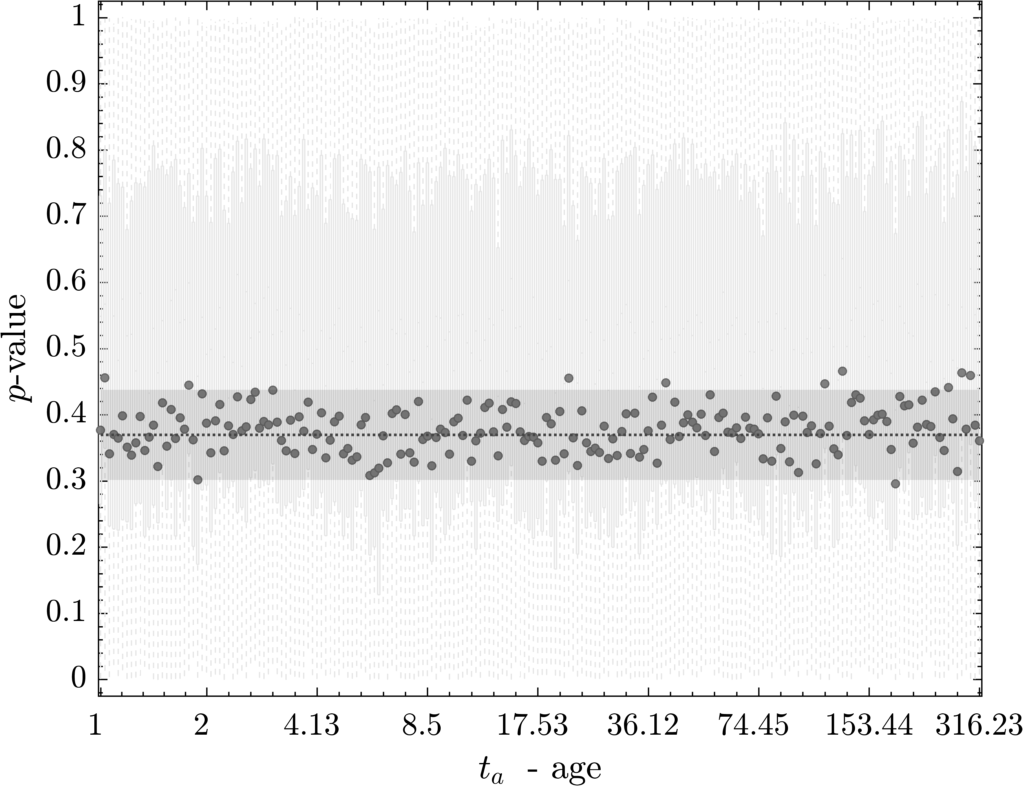}
       	                  \caption{ Power Law distributed waiting times for events with $\mu=1.5$.}
       	                  \label{}
       \end{subfigure}
       \caption{}
        \label{fig_powers2}
 \end{figure}

Other then poisson processes with exponential inter-events time distributions, we now consider non-poisson renewal processes with power-law inter-events time distribution. 
For this purpose we could use the Manneville map approach \citep{aquino2001sporadic} which produces  renewal events whose waiting times are distributed exactly as in eq.\eqref{renewalprescription} as a Pareto-like distribution, we show in Fig.\ref{fig_powers2} the results of the \textit{XA} test applied to events distributed with $\mu=2.1$ and $\mu=1.5$. Clearly the test reveal no significant memory between the events, so accpeting the renewal hypothesis of the underlying process even when the distribution does not have some or any finite moments.

Also in this case it is important to address the domain for $t_a$, but as regard with  the case of the power law coefficient $1<\mu<2$ we do not have a finite mean-time of the waiting times, so we should always check the l0w-sample situation numerically.

\subsection{Synthetic Non-Renewal Processes}
 We will generate surrogate sequences with a  marginal distribution of correlated inter-event intervals in order to obtain a surrogate process which is not renewal \citep{farkhooi2009serial}.
A typical history-dependent process can be modeled by an autoregressive AR process within the limits of stationarity and ergodicity conditions \citep{brockwell2013time} and a general form of the autoregressive process with serial dependence up to a finite lag $L$ reads:
\begin{equation}
X_s=c+\beta_1 X_{s-1} + \beta_2 X_{s-2}+\ldots + \beta_L X_{s-L} +\epsilon _s
\end{equation}
where $\epsilon _s$ is assumed to be iid variable with the specific mean and finite variance, $\beta _i$ are
the correlation parameters for each specific lag, and $c$ a constant.
For our purpose of generating a surrogate non-renewal process, we will only take $\beta_1 = \beta \neq 0$ in the stationary case of $|\beta|<1$ so we have the AR(1) process as:
\begin{equation}
X_s=c+\beta X_{s-1} +\epsilon _s
\end{equation}
where $\epsilon _s$ is taken normally distributed with zero mean and unit variance.

 At this point, as an example, we can mimic the inter-arrival time periods in two different ways, a linear transformation of AR model and an exponential transformation of $X_s$ in the case $c=0$. The correlation structure dies off geometrically as the lag increases.

In the first case the inter-event times intervals could be taken as:
\begin{equation}
\Lambda _s= \big|X_s - E(X_s)\big|
\end{equation}
so that the waiting times are positive and $E[\Lambda _s]= \sqrt{2\sigma_X^2/\pi}$ where $\sigma^2_X=\sigma_{\epsilon}^2/(1-\beta ^2)$.

\begin{figure}[!ht]
       \centering
      	 \begin{subfigure}[c]{0.45\textwidth}
      \includegraphics[width=1\linewidth]{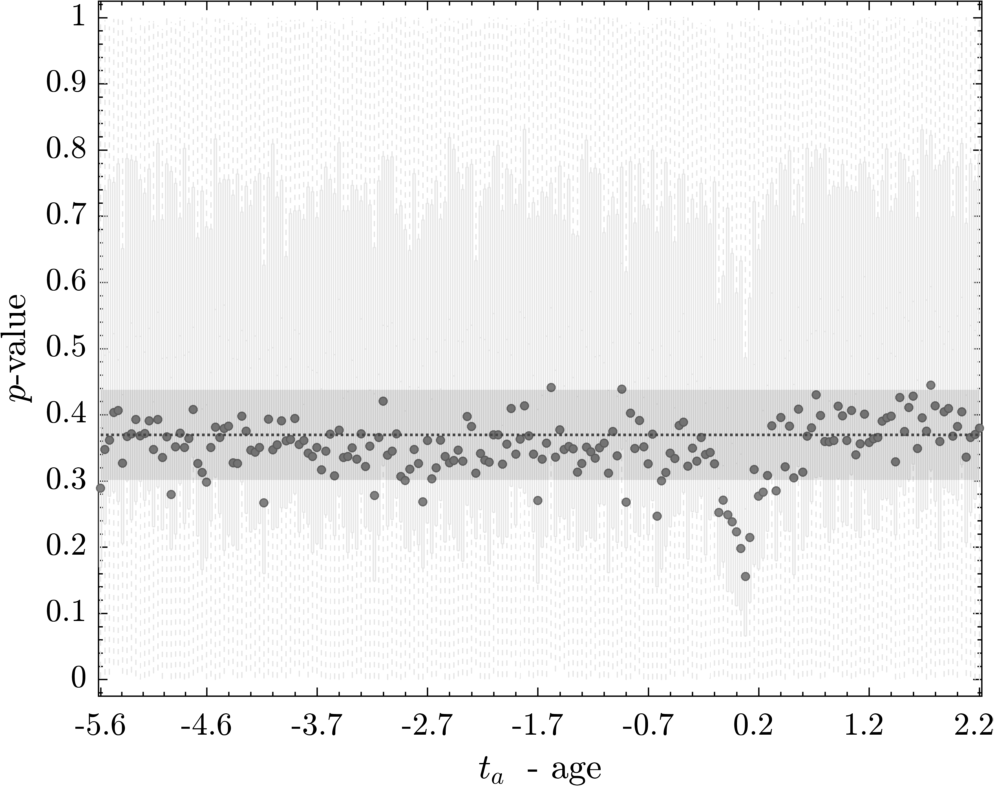}
        	                        \caption{ $\beta=0.3 \rightarrow E[\Lambda] = 0.84$ }\label{}
            \end{subfigure}%      
               %\hspace{2\textwidth}
       	       \vspace{20pt}
       	       \qquad
  \begin{subfigure}[l]{0.45\textwidth}
       	 \centering
       	                  \includegraphics[width=1\linewidth]{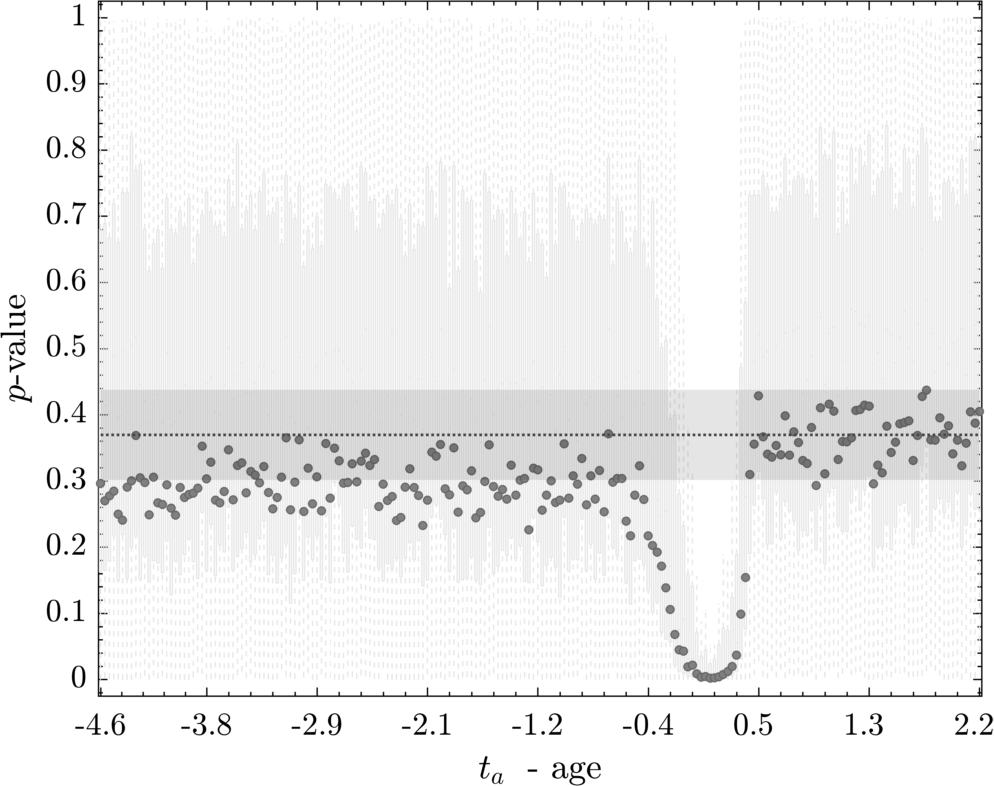}
       	                  \caption{ $\beta=0.5  \rightarrow E[\Lambda] = 0.92$}
       	                  \label{}
       \end{subfigure}
       \\
       	 \begin{subfigure}[c]{0.45\textwidth}
             \includegraphics[width=1\linewidth]{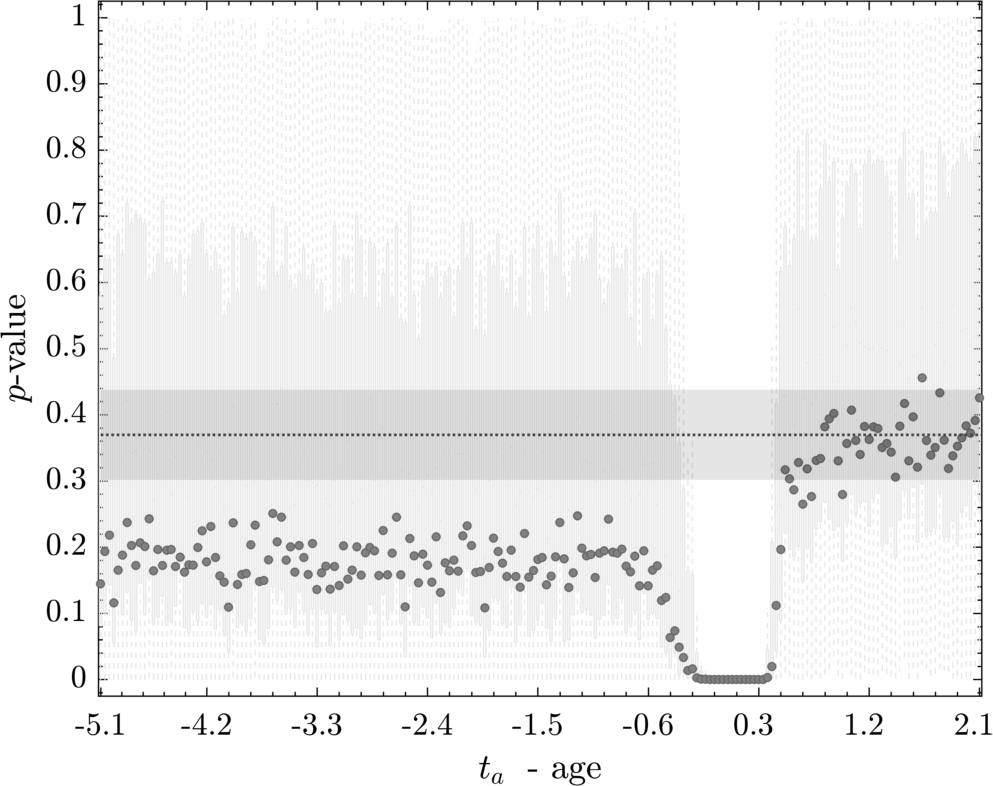}
               	                        \caption{ $\beta=0.7  \rightarrow E[\Lambda] = 1.12$ }\label{}
      \end{subfigure}
      \qquad
             	 \begin{subfigure}[c]{0.45\textwidth}
                   \includegraphics[width=1\linewidth]{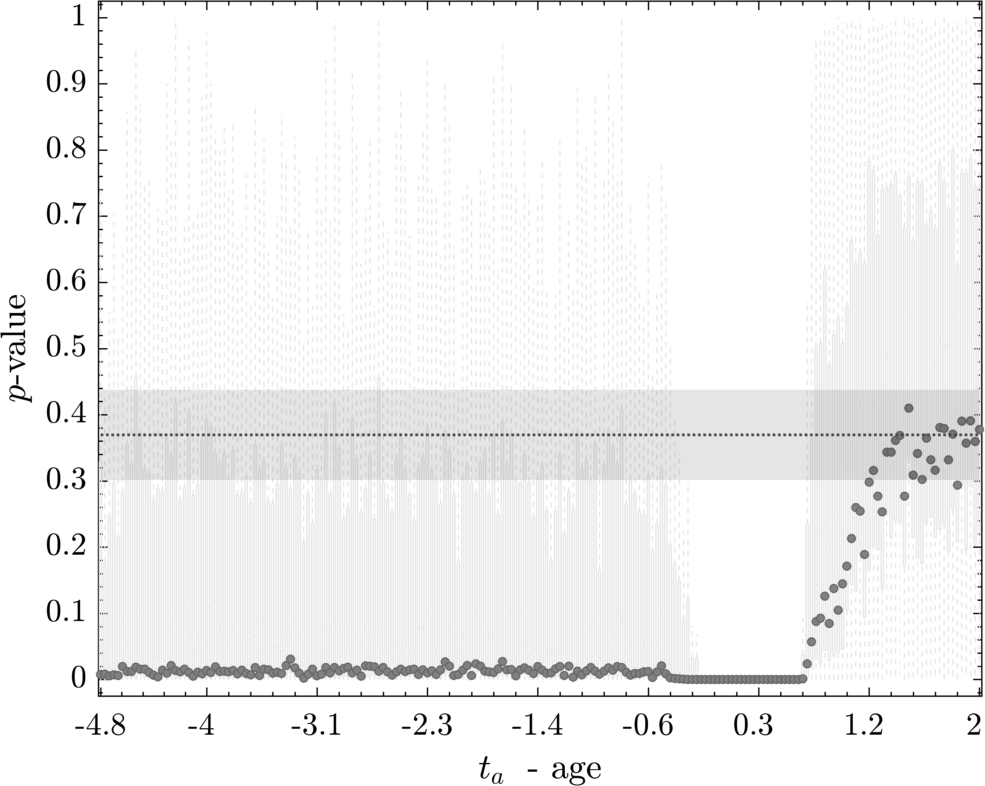}
                     	                        \caption{ $\beta=0.9  \rightarrow E[\Lambda] = 1.83$ }\label{}
            \end{subfigure}
       \caption{}
        \label{fig_absAR}
 \end{figure}

The  \textit{XA} plots applied to linear auto-regressive waiting times is then plotted in Fig,\ref{fig_absAR}, where it is possible clearly see that

In  the other case of exponential transformation \citep{granger1976forecasting}, we define the inter-arrivals time intervals as:
\begin{equation}
\Delta _s = e^{X_s}=e^{\beta X_{s-1} +\epsilon _s}
\end{equation}
where $\Delta _s$ is the series  of correlated intervals, $\beta$ describes the negative serial dependence of the series $X_s$  and $\epsilon _s$ is an iid normal variable with zero mean and unit variance. 
The resulting log-normal distribution of $\Delta$ has mean and variance as:
\begin{align}
E[\Delta_s] & = e^{\frac{1}{2(1-\beta ^2)}} \\
\text{Var}[\Delta _s] & = e^{\frac{1}{2(1-\beta ^2)}}  \left( e^{\frac{1}{2(1-\beta ^2)}}  -1 \right)
\end{align}
In such process the rate of events is $\lambda =1/E[\Delta_s]$ so the maximum ages where we can perform the test is $t_a^{max} = \lambda T/30$ under the condition $\lambda > 1/\sqrt{e}$. Let us now perfor our renewal statistical test on such kind of non renewal test, in the specific choice of the AR model parameter of $\beta = 0.674$ so that $\lambda = 0.4$, and a simulation length of $T=10^4$, we obtain a $t_a^{max}$, in Fig.\ref{fig_ARno} we recover the evidence against the null hypothesis that the process is renewal since all the ages, the geometric mean points are always outside and below the confidence stripe of the null hypothesis: this confirms the presence of intense memory in the event process.

 \begin{figure}[!ht]
       \centering
      	 \begin{subfigure}[c]{0.45\textwidth}
      \includegraphics[width=1\linewidth]{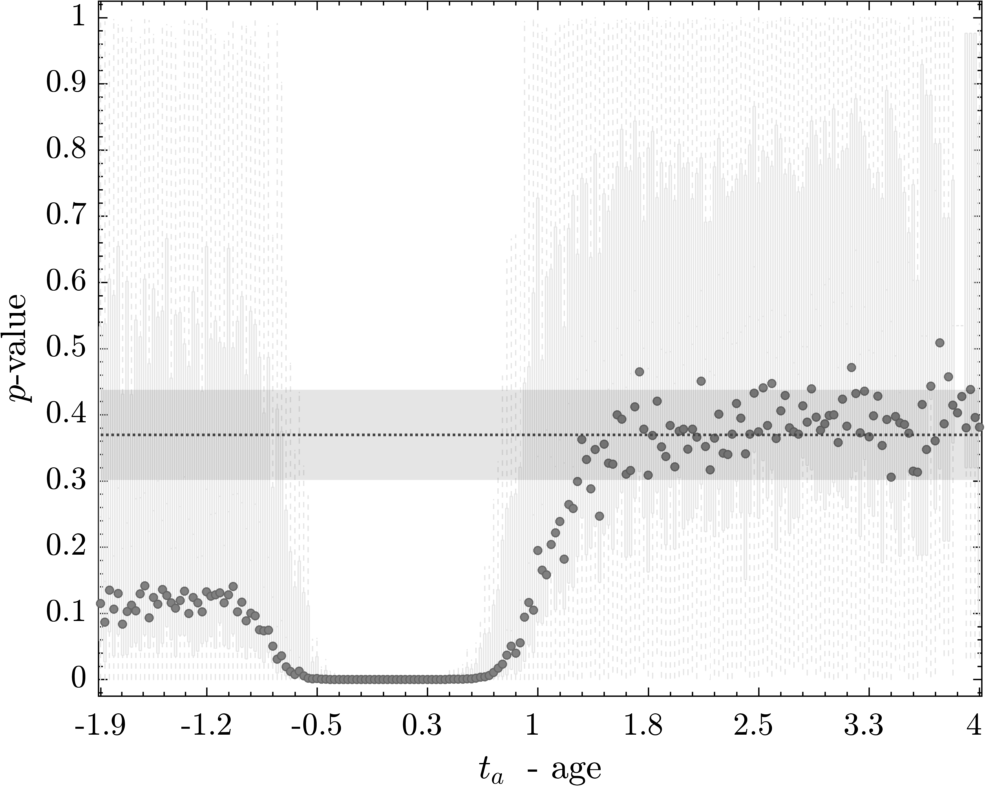}
        	                        \caption{$\lambda = 0.5$, $\beta=0.53$  }\label{}
            \end{subfigure}%      
               %\hspace{2\textwidth}
       	       %\vspace{20pt}
       	       \qquad
  \begin{subfigure}[l]{0.45\textwidth}
       	 \centering
       	                  \includegraphics[width=1\linewidth]{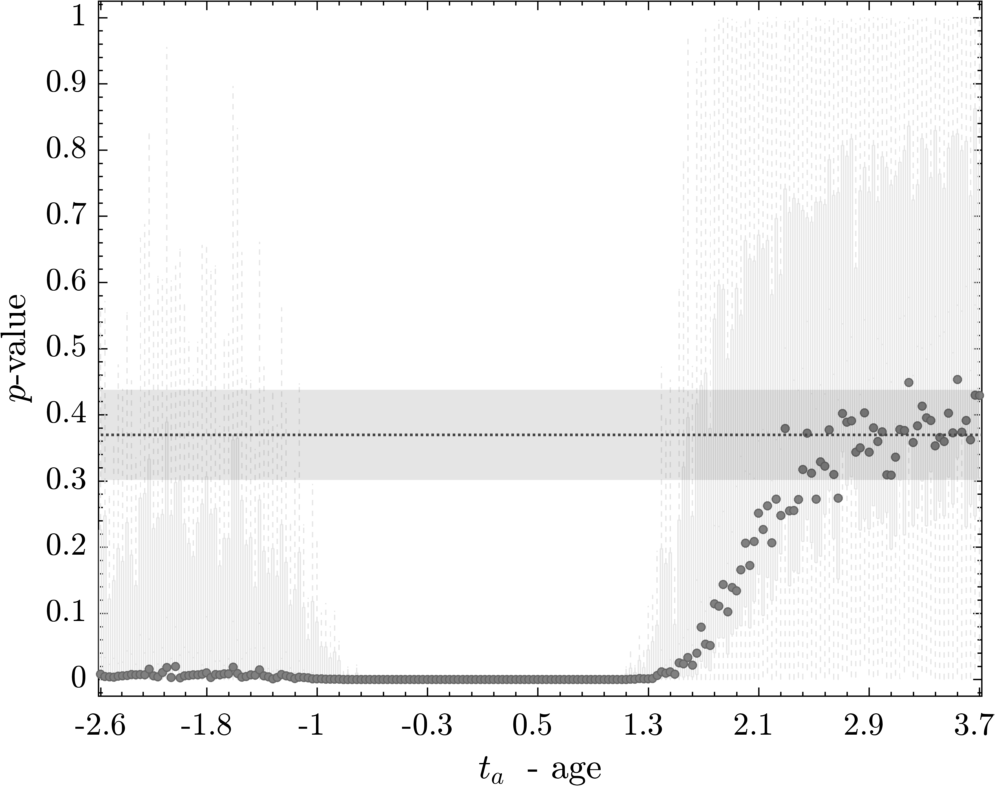}
       	                  \caption{ $\lambda = 0.25$, $\beta=0.8$}
       	                  \label{}
       \end{subfigure}
       \caption{Surrogate not-Renewal exponential AR process  so that the rate of events is  $\lambda = 0.25$. The renewal  hypothesis testing via aging experiment reject the renewal assumption as an overall result of the test.The gray circles are the geometric means for each age, and they clearly do not fluctuate around the expected  value (dashed line) under the null hypothesis.}
        \label{fig_expAR}
 \end{figure}

At this point, we apply our renewal test to a class of non-homogeneous poisson process where the instantaneous event rate is modulated by the past occurrence of events, breaking any renewal property in the process. In particular, among the possible self-exciting models. we select an Hawkes process with exponential kernel \footnote{We produces simulations of  Hawkes processes using the Ogata's thinning algorithm \citep{ogata1981lewis} as described in \citet[ch. 7.5]{vere2003introduction} using a modification of the simulation toolkit in the package described in \cite{xu2017thap}.} so that the event rate is defined as:
\begin{equation}
\lambda (t) = \lambda _0 + \sum_{j:t_j<t} \alpha e^{-\beta (t-t_j)}
\end{equation}
so that each arrival of an event in the system increases the arrival intensity by the factor $\alpha$, after the event, the arrival's influence decays at rate $\beta$.

\begin{figure}[!ht]
   	 \centering
   	  	                  \includegraphics[width=0.6\linewidth]{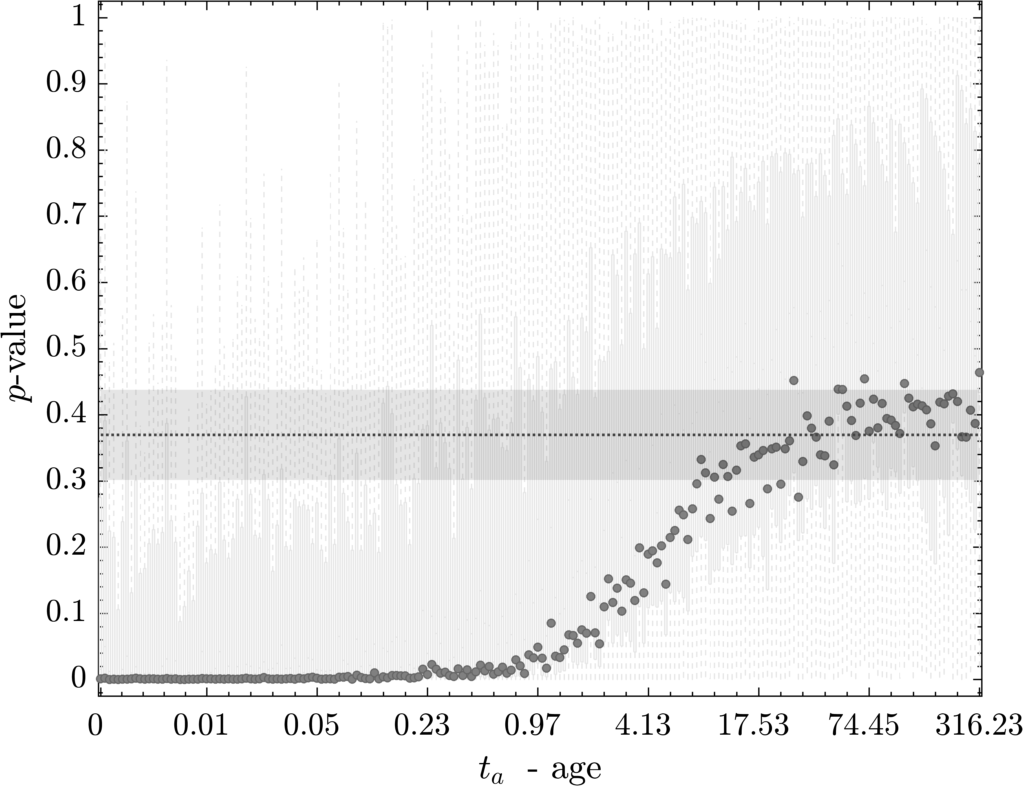}
   	                        \caption{Surrogate of a stationary  Hawkes process with exponential kernel.  $T=4e3$ and $\lambda_0 = 0.75, \,\alpha=0.2,\, \beta=0.4$. We can observe that the geometric mean points have a transition from a intense memory to a renewal condition passing through the the values from which the exponential kernel decays.}
   	                        \label{fig_HW1} 
   \end{figure}

The process is stationary if  $\alpha<\beta$ and we have that $\overline{\lambda}=\frac{\beta}{\beta-\alpha}\lambda_0$ is the average rate of events\footnote{Notice that in the case of $\alpha >\beta$, no mean rate ($\overline{\lambda}$) of events is defined, and we should use the same procedure as in the power-law inter-event case, where we numerically check  the low-sample condition.}.
Choosing $T=4e3$ and $\lambda_0 = 0.75, \,\alpha=0.2,\, \beta=0.4$, the maximum age we have $t_a^{max}= \overline{\lambda}T/30 \approx 10^2 $. 
We plot in \ref{fig_HW1} the \textit{XA} test, in  which we can see two an initial not-renewal feature of the system for short ages up to the order of the exponential decay in the memory kernel of Hawkes process.  While as global test one have to reject the renewal assumption, the $XA$ plot allows to check the renewal conditions a different temporal scales: in this cases, a short time scale we detect memory between events, but, after a transition, we see how at large time scale the events looks without any memory.

\subsection{Superposition of events}

The case of the Hawkes process described above, is a typical case of spurious process with mixed behavior with renewal and not renewal patterns at different time scales.

However, one can go beyond a single process which produces events of different memory scales. One can in principle have a series of events generated by different underlying processes.  There are, in fact, many ways to produce generalized renewal processes \citep[ch.9]{cox1965the}, for our purpose we select the specific case of processes' superposition \citep{cox1954superposition, cinlar1968superposition,teresalam1991superposition}. It consists in considering the case where there is a number of independent sources at each of which events occur from time to  time. 

Let $\{A_n\}$ and $\{B_n\}$ be two independent point processes in general with different interrenewal distributions. The pooled process has a number of  events as:
$$
N(t):=N_A(t) + N_B(t)=\sum_{n=1}^\infty \left[\mathsf 1_{(0,t]}(A_n) +\mathsf 1_{(0,t]}(B_n)\right].
$$
 In general, the correspondent point process is renewal for only particular situations \citep{ferreira2000pairs}. For example  the superposition of two poisson renewal processes produce a pooled process that is renewal and with the sum of the rates of the original exponential inter-arrival distributions. 

\begin{figure}[!ht]
   	 \centering
   	  	                  \includegraphics[width=0.65\linewidth]{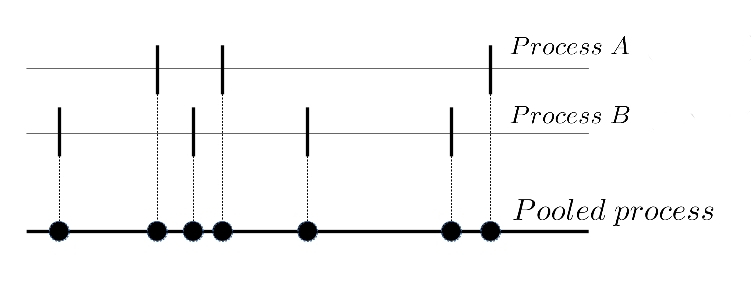}
   	                        \caption{ Superposition of two independent sources of events. The process $A$ generates renewal events, the process $B$ generates not-renewal events. The pooled process is a spuriou sequence of events .}
   	                        \label{fig_pool} 
   \end{figure}

We take a particular case in order to produce a given pattern of events' sequence  not present in the previous examples. For that reason, let us take two independent point processes: one  is renewal and the other is not renewal, the resulting superposition of the twos, as show in Fig.\ref{fig_pool}  is a process formed by pooling  the two types of events.

In particular, let us take process $A$ as a renewal poisson process and the process $B$ as a not renewal process (as in the auto-regressive case). We, in particular, consider the case when the rates of the two sources of events are $\lambda_A>\lambda_B$ and in a particular case the  renewal process has a rate $\lambda_A=8$ and $\lambda_B=0.75$, so the two different scale are at least one order of magnitude different.

The results of $XA$ test is shown in Fig.\ref{fig_XApool} where it is clear that, or that particular case, the XA plots show a clear two time-scales at which the process have events without memory at small ages, and instead it shows renewal events for larger ages.  The transition in memory happens ate ages closer to the average inter-arrival times of the not-renewal processes where its rate dominate the higher rates renewal events.  However it is important to notice that in general we cannot infer from XA plots the properties  original sources of the pooled sequence. One can have many types of superpositions. The only inference one could make using the \textit{XA} test is to possibly detecting time-scales for which the events show memory. Further, the response of the test as in Fig.\ref{fig_pool} 

 \begin{figure}[!ht]
       \centering
      	 \begin{subfigure}[c]{0.475\textwidth}
      \includegraphics[width=1\linewidth]{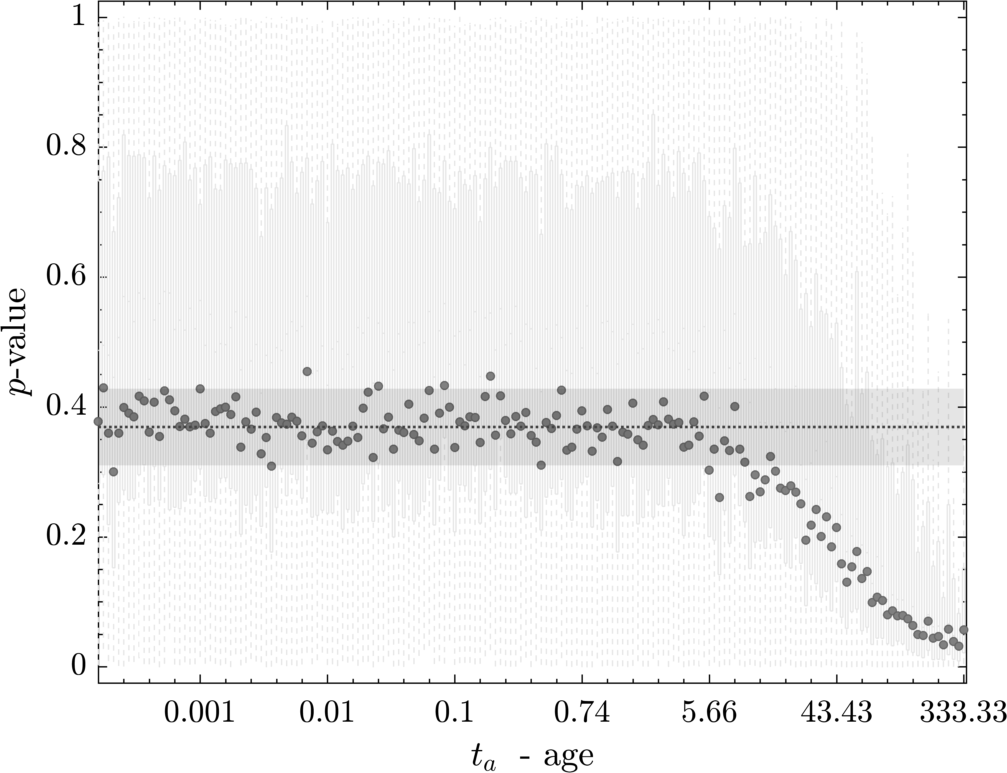}
        	                        \caption{Sequence of pooled events with $\lambda_A > \lambda_B$}\label{}
            \end{subfigure}%      
               %\hspace{2\textwidth}
       	       %\vspace{20pt}
       	       \qquad
  \begin{subfigure}[l]{0.475\textwidth}
       	 \centering
       	                  \includegraphics[width=1\linewidth]{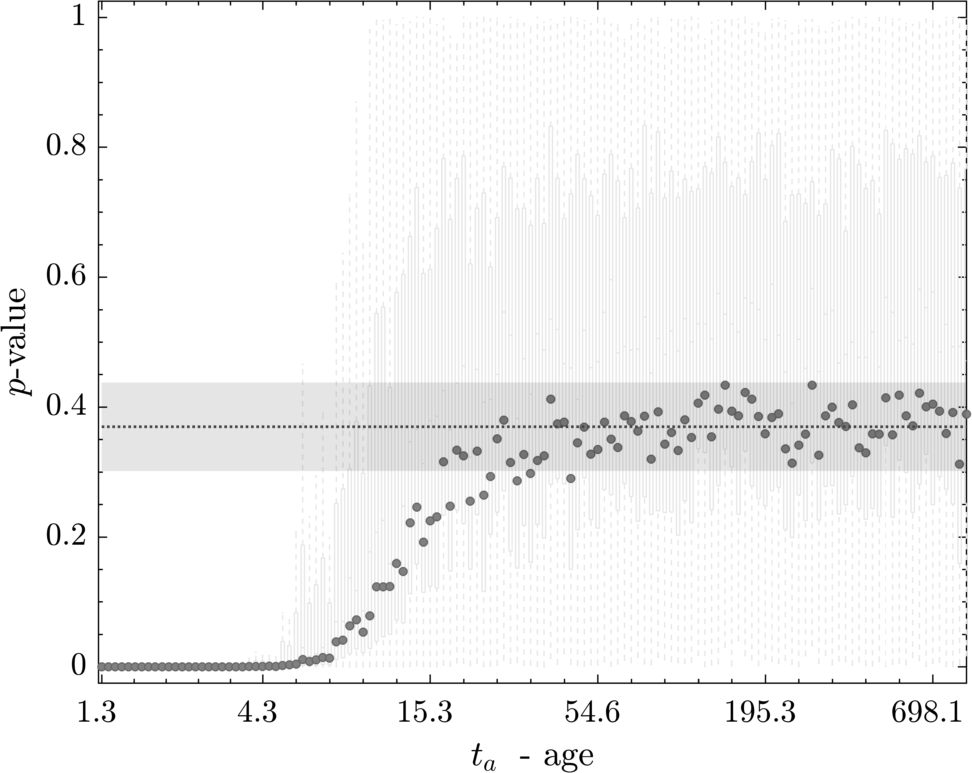}
       	                  \caption{Sequence of pooled events with $\lambda_A < \lambda_B$}
       	                  \label{}
       \end{subfigure}
       \caption{ Superposition of two independent sources of events. The process $A$ generates renewal events, the process $B$ generates not-renewal events. The pooled process is a spurious sequence of events. In the case (a) the renewal process generates events with average inter-event intervals $\langle \tau \rangle_A=1/\lambda_A =0.125 $ and the not-renewal events have inter-arrival intervals $\langle \tau \rangle_B=1/\lambda_B =1.\bar{3}$. In the case (b)  the renewal process generates events with average inter-event intervals $\langle \tau \rangle_A=1/\lambda_A =0.5 $ and the not-renewal events have a mean inter-arrival time interval $\langle \tau \rangle_B=1/\lambda_B =0.125$.}
        \label{fig_XApool}
 \end{figure}

Superposition of point processes is an important class of stochastic processes for its wide range of possible applications where sequences of activities arrive at a central collector of events from a number of independent sources. For example,  in production networks, during an industrial stage, several machines operate independently in parallel. The sequence of times at which items are produced follow a superposition of processes. In managing the next stage of production one can  find useful to know the properties of the pooled sequence of events. Another application of a superposition of renewal processes is used to model the effect of imperfect maintenance \citep{kallen2010superposition} and arrival processes applied to model queue behaviors \citep{albin1984approximating}.

\section{Single sequence of events: approximated \textit{XA} test }
In many practical case, the researcher has only single (or few) observation of the process, in that case our exact renewal test is not applicable since it requires the assumption of  many independent samples. In the worst case scenario, one have data with a single observation of the events and consequently only one realization of the point-process time series.

However, in such situation, it is possible to set up a statistical technique based on the available data which extend the use of our exact \textit{XA} test to an approximated test valid for single realizations of events' sequences.

In the case that the data consist only in one sequence it can be considered as a single realization of the process generating the events. Moreover, we also consider the worst case scenario where the observation is made up of few events implying so  a low statistics of the sample. The most crucial statistical problem in applying the exact $XA$ test  is that one has to guarantee  the independence among each pair of samples in the two-sample confide nce test (for example in the Kolmogorov-Smirnov test).

In this case, we propose  two combined resampling  techniques which try to minimize the dependence in the data due to the fact that the hypothesis test has to be performed on a single observation of the process.

Essentially, we use split the original sequence making as many as independent randomized sub-samples from to original realization and then perform two-samples test without using the K-S test.

There are several popular resampling techniques which are often used in computational statistics and machine learning \citep{good2004permutation} and which can be used to build an approximated  version of the the \textit{XA} test. We have focuses our study on two resampling methods for single observation test: one method (\textit{bootstrapping}) is used to create independent samples and another randomization technique (\textit{permutation test}) will be used to perform the statistical test  replacing the Kolmogorov-Smirnov test which suffers of small, not independent and discrete samples. 
The only assumption made up by resampling approaches  is that the observed  data are a representative sample from the underlying population but no assumptions are made on the population distribution and parameters. 

The difference between the exact \textit{XA} test and its approximated version is focused on the two-sample significance test as in Fig.\ref{fig_scheme}. Since we cannot generate other sequences of the inter-arrival times under the null hypothesis we can infer the behavior of the population by the only observation we have, bootstrapping the distribution of the inter-events times from the observed events, so obtaining estimates of \textit{p}-values from the two-samples tests and finally the geometric means variable $\tilde{g}_p(t_a)$.

\paragraph{Surrogate events from  Bootstrapping}
Let us call $L$ the data size of the sample set (i.e. the length of  events' time series), $\{\tau_1, \tau_2,\ldots,\tau_L\}$. We split the  sample set in $t_w$ not-overlapping windows to create replica of the process with sub-samplings of the original dataset. In such way, $ N=L/t_w$ sub-samples are considered as independent realizations of the waiting times's sequence of events. Since we  use the two-sample significance test,  we need  independent samples for the shuffled (no-memory) waiting times.  Without knowledge  of real distribution of the observed $\tau _i$'s,  we use the empirical cumulative density function $F_n(\tau)$ as an estimate of the original cdf $F(\tau)$.  Subsequently, we sample from  empirical density function using a random generator of events which is equivalent to sampling with replacement from originally observed sequence.  Sampling from $F_n (\tau)$ is equivalent to sampling with replacement from originally observed inter-events times.
 The aging experiments and the two sample tests will be computed comparing the samples in  one of the $N$ windows with other samples of the same size $t_w$ drawn from the bootstrapped distribution. In this way we can guarantee independence among the two distributions in the two-sample test, i.e. the samples within scrolling windows versus the bootstrapped samples.\footnote{In alternative we also used $k$-fold Cross-Validation resampling technique: the observed dataset is partitioned into $k$ groups, where each group is given the opportunity of being used as a held out test set leaving the remaining groups as the training set. As results the bootstrapping technique produce a more extended age $t_a$ allowing larger time-scales to be explored.}. 

\paragraph{Two-sample Permutation tests}
As  statistical significance test the distribution of the test statistic $D_{m,n}$ in eq.\eqref{eq_teststat} under the null hypothesis is obtained by calculating all possible 
values of the test statistic under rearrangements of the 
labels on the observed data points.
In the specific of the two-sample problem we will replace the non-parametric Kolmogorov-Smirnov test  with a permutation test which can be 
used without any assumptions on the  distribution of data. In fact, a permutation test gives a simple way to compute the sampling distribution for any test statistic under the null hypothesis. The statistical significance of the permutation test, as expressed in a $p$-value in eq.\eqref{eq_pvalue}, is calculated as the fraction of permutation values that are at least as extreme as the original statistic, which was derived from non-permuted data.   If  the  null  hypothesis  is  true  the  shuffled (randomized) data  sets  should  look like  the  observed  data (within the time windows),  otherwise  they  should  look  different  from  the real  data. Despite that the permutation test is an exact test, it usually could be extremely costly in terms of computational resources. In particular, choosing the same number of samples in each sequence, there are exactly $s=\binom{2t_w}{t_w}$ ways of randomly allocating $t_w$ of the observed time-intervals and the remaining $t_w$ of the bootstrapped intervals.
Since the exact permutation test can be computationally intensive, we also allow the use of an empirical method directly couples both the minimal obtainable $p$-value and the resolution of the $p$-value to the number of permutations. Thereby, one can impose a maximum number of permutations, i.e. $s=1000$, we have that $p= 1/s=0.001$ is the smallest possible $p$-value. However if the length $L$ of the single realization of the events is very large the sample-size can be enough to perform the Komlogorov-Smirnov test or any parametric two-sample test instead of the very expensive permutation test.

\paragraph{Single realization XA plots}
The approximated $XA$ plot for single realization is obtained in the same way as in the exact $XA$ test, expect that in Fig.\ref{fig_scheme} one should take $A_i$ as sample of size $t_w$ in a given time interval of the entire sequence, and the sample $B_i$ is the sequence of events generated by the bootstrapped distribution. Moreover the two-sample test could be performed using a permutation test other than the non parmametric Kolmogorov-Smirnov test.

Another important difference in performing the approximated $XA$ test is given by the presence of the new parameter $t_w$ which determines a series of consequences summarized in Table \ref{tab_par2};  the fact that we split the entire unique sequence in many pieces introduces more constraints in the test which are not present in the exact $XA$ test as described in the caption.
\begin{table}[!ht]
\centering
\begin{tabular}{|c|c|c|c|c|}
\hline
 $t_w$  & $t_a^{min}$        & $t_a^{max}$      & $T_a$&  $N$\\ \hline \hline
   
 \textit{accuracy} & \multicolumn{1}{c|}{$ \min\{\tau_i\}$ } &  $\; \sim \min\{L / \langle \tau \rangle \,,\, t_w \}$ &  $\lesssim 1/ \min\{\tau_i\}$  & \textit{dof}  $\approx \, L/t_w$ \\ \hline
 
       &         smallest    &   largest    &  \textit{temporal resolution}    &   \\ % \hline
         &        memory block     &   memory block      & test's sample size  & statistical precision \\ \hline
\end{tabular}
\caption{Parameters of the single realization \textit{XA} test. One one have a single observation, $t_w$ is a new parameter which derives from the fact that we split the unique observed sequence in non-overlapping  intervals and $t_w$  indicates the number of events per window $t_w=L/N\gg 1$. Then,  $T_a$ is the number of $t_a$'s used in the repeated tests representing the sample size  of geometric means in the whole statistical procedure.  But in the case of single realization \textit{XA} test the step-size $\delta _{t_a} =(t_a^{max}-t_a^{min} )/T_a $ should be chosen in such a way to keep the aged-sequences independent so $\delta _{t_a}\gg  \min\{\tau_i\}$. On the other side, also $N$ is  constrained to the data sample size $L$ as trade-off between precision of the test when $N$ is large and the maximum time scale we can explore $t_a^{max}$ which is related to  our choice of $t_w$. When one increases $t_w$ it is possible to explore the memory on a larger time scale but with the effect of poor precision since $N$ becomes smaller. In the other way around, increasing the precision mean reducing $t_w$ and so the time-scale of the memory blocks.  } \label{tab_par2}
\end{table}

The main limitations of the approximated \textit{XA} test for single realizations, are due to a short range of the time-scale one could explore, low maximal age $t_a^{max}$ and a less power of the test related mostly to the limited spatial resolution $N$  which determines the precision of our confidence about memory in the process.

At this purpose, we re-compute some of the synthetic case in the exact \textit{XA} test in  the single-realization case. 
The approximated \textit{XA} plots in Fig.\ref{fg_dd} clearly shows the ability of the test in detecting memory in the synthetic sequence of  inter-arrival times.

As last resort of correlated events, one should take in account which the meta-observation we reconstruct are not fully independent for many reasons, it is not correct use the Fisher's approach to multiple comparison of p-value. In such situation all the tests have something in common and are considered as a family of tests In such a case the adjustment methods try to ensure that the chances of a Type I error are maintained below the claimed size of the test. In such a corrected Bonferroni method of multiple p-values  will be more reliable since the method will not claim significance unless some individual tests do.

\section{Conclusions}

The main advantage is that one does not have to worry about  distributional  assumptions  of  classical  testing  procedures;  the  disadvantage  is
the   amount   of   computer   time   required   to   actually   perform   a   large   number   of permutations,  each  one  being  followed  by  re-computation  of  the  test  statistic.  Despite In terms of future applications on supercomputers and high performance computing, the combined use of speed processors , parallel techniques and GPU accelerations   would  allow  users  to  perform  any  computational statistical  test
using the permutation method. 

%\section*{Acknowledgment}

% \newpage
% \begin{appendices}
% 
%
% \end{appendices}
% 
 %\newpage
 %\theendnotes

%\include{appendici\appendici.tex} 
 
 %\addcontentsline{toc}{section}{Bibliography} 
%\bibliographystyle{authordate1} 
% \bibliographystyle{apalike}
 %\bibliographystyle{plain} 
 \section*{References}
 \bibliographystyle{apalike}
 \bibliography{biblioR2} 
 
% \appendix
 %\input{appendici.tex}

 \section*{A review of renewal processes }
 
 In the recent literature of complex systems made up of interacting agents, the presence of bursty activity  in temporal evolution is revealed by a certain waiting time distribution of consecutive events \citep{karsai2018bursty}. If the distribution of those events has a power-law form, such kind of events are expected to produce aging effects in the corresponding time-integrated network \citep{moinet2015burstiness}. Moreover such burst patterns can be at the level of single individuals  or at the level of the whole system, and this can have important impacts on the dynamics of spreading processes \citep{min2011spreading},  \citep[pg161-174]{holme2013temporal} both in terms of decision making and consensus and in terms of contagion and response to shocks.

 It is of a crucial interest to know if a model or real world data shows such renewal patterns in its event activity, in order to select the correct model or to give a correct interpretation of real world systems. For example models based on renewal assumptions as the Continuous Time random walk framework \citep{scalas2006complex,namatame2006complex} or on Fractional Brownian motion  obtained  as  the  limit  of  a  superposition  of renewal   reward   processes   with   inter-renewal   times   with infinite variance \citep{levy2000renewal}.

 \subsection*{Renewal patterns in  Economics, Finance and Natural Sciences}
 There are many areas of economics which make use of renewal theory even if usually such framework is not fully taken in account. In order to give a more concise view  of renewal theory, we will make a short review of some topics where such theory has been employed and we will discuss the more recent studies where detection of memory between events and renewal property could be useful.
 
 Historically, the concept of renewal processes arises from  mathematical, physical sciences and engineering  This type of analysis is characteristic of the applications of renewal theory to areas such as population dynamics, the theory of collective insurance risk, and to the economic theory or replacement and depreciation.

 Renewal theory plays an important role in the field of financial time series such as stock prices, foreign exchange rates, market indices and commodity prices. In particular, there is an unsolved debate about which kind of mechanism brings all the well-known stylized facts financial time series analysis regarding, among the others, the presence of long memory properties, fat tails returns distribution and volatility clustering, as persistence of the amplitudes of price change \citep{cont2005long,cont2007volatility,mantegna2007introduction}.  
 In the economic literature there are examples where agents switch  between two behavioral patterns which leads  to large aggregate fluctuations. In the  context of financial markets, these behavioral patterns could be some trading rules and the resulting aggregate fluctuations  large movements in the market price. Anyway, ordinary Markov switching models, despite the fact that they can mimic the volatility clustering property, they are not able to explain long-range correlations i.e.  the time spent in each regime –the duration of regimes– is not heavy tailed distributed.  By contrast with Markov switching, which leads to short range correlations,  renewal switching models have been introduced \citep{leipus2005renewal} \citep[ch.3]{teyssiere2006long}. in order to match with the  strong empirical evidences of heavy tailed  correlations. For example,   the daily S$\&$P composite price index exhibits long memory in volatility and heavy tails  \citep{liu2000modeling}; furthermore  in the length of the US business cycles \citep{jensen2006long} the timing of the boom and bust states constitutes a renewal process.

 Also regarding with bank liquidity management \citep{cabello2013cash}, renewal theory can capture the random elements of the cash flow,  within the major issue of financial crisis as liquidity shortages. Those kind of models try to assess the  bank branch cash management, focusing on the optimization of cash inventories as a  critical feature of financial intermediation.
 
 Another important field in economics is the present value analysis \citep{washburn1992present} with the presence of renewals events in situations where a time sequence of (net) cash flows  would be predictable except for the  presence of occasional renewals that force the cash flow to begin anew after each one.
 
 Renewal theory has been used in the field of the management science, financial risk and structural safety as regard with investment and maintenance optimization  in finding an optimal balance between the initial cost of investment and the future cost of maintenance. This framework can be  modeled as a renewal process if one can identify independent renewals that bring a system or structure back into its original condition. As regard with the costs of such decisions,  some authors \citep{van2003explicit, van2008renewal} take in account  the time value of money by discounting and to consider the uncertainties involved with costs that can be discounted according to any discount function such as exponential, hyperbolic and no discounting.
 
 In particular, in the theory of financial risk management for natural disasters, the theory of renewal process has been used to model and provide an index of the monetary damage from such disasters as earthquakes, wind storms, floods and other natural disasters \citep{batabyal2001aspects, pandey2017stochastic}, where renewal theory is also used to model a large class of natural resource regulatory problems involving systemic and policy uncertainty,  which involves decision-making  in a dynamic and stochastic environment \citep{batabyal1994renewal}. 
 In the probabilistic modeling of life-cycle management, the renewal theory plays a key role in the computation of the expected number of renewals and the cost rate associated with a management strategy; in particular, in an ecological-economic perspective a system's cycle over time and its dynamics consist of shocks which occur in accordance  with a renewal process and it generates a set of ecological and a set of economic effects \citep{mitov2014renewal}.  This  clearly indicates that this problem of an ecologically unsustainable world arising from the dichotomy between innovation and environment is the central issue of climate change debates in the last few decades. For example, \citet[ch.4]{batabyal2008dynamic} faces the decision to use or not fertilizers to enhance or overseeing the problem of soil fertility deterioration using a theoretical model  based on the renewal processes theory. In such way the authors identify some agricultural harvest  policies to calculate cost-reward of  decisions about the management of the problem of soil fertility.

 Renewal processes plays also a key role in economic modeling as for example in the work of \citep{chipman1977renewal} where  a model of economic growth is designed to provide a formal framework  dealing with the problem of optimal selection of investment projects. 
 
 %\subsection{Renewal patterns in Biology and Neuroscience}
 On the side of natural sciences, renewal models has been widely used  to describe  some of the phenomena of genetic linkage  and chromosome maps where one assumes a renewal property for the points of exchange that occur on a single chromosome strand during  the appropriate  stage of mieosis \citep{bailey1961introduction,bailey1990elements,speed2012genetic,lange2003mathematical}.  Another important application of renewal theory is about properties of ion channels, i.e. structures contained in membranes of cells, such as those found in heart and nerve tissue. Such renewal-theoretic approach is also  useful if Markovian models are not appropriate as shown by \citet{dabrowski1990renewal}.
 
 The spontaneous activity of a neuron as well as its response to repeated stimuli
 is  often  characterized  by  irregular  and  unpredictable spike  trains as consequences of various  sources  of  neuronal  noise.  Simplified stochastic models have been suggested,
 in order to assess the effect of fluctuations and most of these models generate spike trains with independent interspike intervals (ISIs) within a renewal point processes framework. However, recent studies \citep{avila2011nonrenewal,lindner2004interspike} have provided experimental evidence for non-renewal spiking, reporting significantly large  correlations between ISIs for various types of neurons.
 
 For the sake of completeness regarding to the bursty activity and renewal patterns of systems, we mention another important type of counting process called self-exciting processes since the intensity of the process is driven by a function of the recurrence time to all previous points. In this way, the intensity is high whenever we have observed many events in recent periods. Such  processes and in particulare the Hawkes ones \citep{hawkes2018hawkes}, naturally accounts for events which are clustered in time and is well suited to model, in finance, the evolution of market activity and trading intensities. They are able to  capture event clustering and thus positive autocorrelations in event durations  and they may account  for the dynamics of market prices at microstructural level \citep{bacry2014hawkes}. In summery, Hawkes process are an important generalization of non-homogeneous poisson processes and they  can be roughly represented and generated by clusters of Poisson processes \citep{hawkes1974cluster}. Hawks type processes have their main empirical applications to address many different problems in high-frequency finance so describing the assets' prices dynamics,  estimating the market stability and accounting for systemic risk contagion among many other applications \citep{bacry2015hawkes,bormetti2015modelling,schneider2018modelling, khashanah2018slightly}. For our purposes, it has been important to have mentioned the Hawkes processes because they are examples of models showing bursty and clustered activities without a necessary renewal patterns (up to some trivial situations). 
 The insight about the presence of renewal property of the underlying process can be crucial in selecting the right model and the correct interpretation of data outputs.  As regarding with the study of earthquakes' recurrence times \citep{saichev2007theory}, it is evident that earthquakes  show both renewal and not renewal components in its evolution according to different hidden characterizations involved in the phenomena. In particular,  various point-process based models can be applied to describe such bursty phenomena  \citep{ogata1988statistical}, some models are founded on renewal and recurrence theory  \citep{garavaglia2010renewal,akimoto2010characterization},  others on clustering models based on self-exciting processes \citep{pratiwi2017self,hawkes1973cluster}. There are, in fact, many results about observing different structures for different temporal scales and magnitude scale; large earthquakes (mainshocks) fits better the bursty renewal patterns in its events evolution \citep{mega2003power,sykes2006repeat}, on the other side small episodes (i.e. aftershocks and foreshocks) seem to be more clustered and correlated  which cannot be addressed using a simple renewal in the case it not possible to neglect the dependence on the stressing history and the impact of earthquake interactions assumption \citep{christensen2002unified,bak2002unified,NonPoissonianearthquake}. The implications of such discussion is the necessity of identification indexes for a more efficient earthquake forecasting models \citep{ogata1988statistical,talbi2013interevent}. It is of high relevance to develop statistical tools in order to detect the temporal patterns in inter-event times which could also whcih  be  applied to economics, finance and business as already happened for many other temporal features for long and short range memory issue \citep{ohanissian2008true, stindl2018likelihood}.

 \subsection*{Aging effect in renewal proecsses}
 A stochastic process $N(t)$ that counts the number of some type of events occurring during a time interval  $[0,t]$ is called
 a renewal process, if the time elapsed  between consecutive events are, typically, independent and identically distributed random
 variables. In particular if successive events are separated by scale-free waiting time periods,  the underlying process exhibits aging: events counted initially in a time interval  $[0,t]$  are statistically different from events observed at later times $[t_a , t_a+t]$.

 In particular, let us suppose $0 \leq t_1 \leq t_2 \leq \ldots $are finite random times at which a certain event occurs. 
 The number of the times $t_n$ in the interval $(0, t]$ is:
 \begin{equation}
 N (t) = \sum_{n=1}^\infty  \mathbf{1}(t_n \leq t),\quad t\geq0
 \end{equation}
 we will consider $t_n$ as points (or locations) in $\mathbb{R}^+$ with a certain property, and $N(t)$ is the number of points in $[0, t]$. The process $\{ N(t) : t \geq 0\}$, denoted by $N (t)$, is
 a point process on $\mathbb{R}^+$ . The $t_n$ are its occurrence times (or point locations)\footnote{The point process N (t) is simple if its occurrence times are distinct: $0 < t_1 <
 	t_2 < \cdots$ a.s. (there is at most one occurrence at any instant). }.
 
 A simple point process $N (t)$ is a renewal process if the inter-occurrence times $\tau_n = t_n - t_{n-1} , \text{ for } n \geq 1$, are independent with a common distribution $\psi$ , where $\psi (0) = 0$ and $t_0 = 0.$  
 
 The $t_n$ are called renewal times, and $\tau_n$ are the inter-renewal times (or waiting times), and $N(t)$ is the number of renewal events in $[0,t]$.

 The epoch of the $n$th occurrence is given by the sum:
 \begin{equation}
 S_n=\tau_1+\cdots +\tau_n
 \end{equation}

 The assumption of ergodicity is  of central importance of statistical physics. It is the hypothesis that ensemble and time averaging coincide, thereby making it possible
 to make statistical evaluations based on the observation of the time evolution of  a single systems, when an ensemble of infinitely many identical copies of the same system is not available. For this reason the experimental observation of the fluctuating fluorescence of  single nanocrystals,  \citep{bawendi}, generated a big interest. In fact,
 the  alternance of light and darkness is proven to be a non-ergodic process \citep{bouchaud}. When luminescence is activated a cascade of bright (''on'') and dark (''off'') states ensues. 
 The time duration of these states tends to increase upon time increase, a clear signal of non-stationary behavior, giving the misleading impression that the rules generating fluorescence intermittency changes with time. It is not so. The dynamical process generating this intermittence is renewal and the occurrence of a switch is a sort of rejuvenation effect bringing the system to select the time length of the new state, either ''on'' or ''off'', with the same probabilistic prescription as that adopted by the system when the luminescence process is activated\footnote{  For simplicity sake, without any loss of generality, we make the assumption that the ``on" states and the ``off" states are characterized by the same waiting time distribution density $\psi(t)$. Actually, the literature on blinking quantum dots shows that the inverse power law indexes $\mu$ are slightly different. This property is of no relevance for the problems discussed in this paper. }.  The compatibility between renewal and non-stationary condition  is due to the fact that the  probability of having a switch from the ``on" to the ``off" state, or viceversa, at a time distance $\tau$ from the last switch, denoted as $\psi(\tau)$,  has the inverse power law structure
 \begin{equation}
 \psi(\tau) \propto \frac{1}{\tau^{\mu}},
 \end{equation}
 with $\mu < 2$. It is well known \citep{feller2} that in spite of the fact that the occurrence of an event at a time $t$  implies a total rejuvenation of the system, and the probability of occurrence of another event at a time distance $\tau$ from it, namely at time $t+\tau$, is given by $\psi(\tau)$, the rate of event tends to decrease as $1/\tau^{2-\mu}$. This non-intuitive property implies a subtle recourse to an ideal ensemble experiment, made with many identical systems, all of them with the property that an event occurs at time $t$. If this ensemble of identical systems were observed, the rate of switch events would decrease in time with the  prescription of Feller \citep{feller2}.

 This non-stationary but renewal condition of blinking quantum dots generated a search for an efficient procedure to establish with satisfactory confidence that their non-stationary behavior is not due to dynamical rules changing with time. This important property can be assessed in principle through the theoretical property of aging. If the observation of the process under study is done at a time $t_a$ far from an event, regardless of whether or not further renewal events occur in between, the distribution density of times that we have to wait to see the occurrence of a new events is different from the distribution density that we may record through the observation of many different realizations \citep{aging1,aging2,aging3}. 
  The histogram must be properly normalized so that the aged waiting distribution density generated by this observation
 fits the important condition
 \begin{equation} \label{normalization}
 \int_{0}^{\infty} dt \psi(t) = 1. 
 \end{equation}
 Since the survival probability $\Psi(t)$ is defined by
 \begin{equation}
 \Psi(t) = \int_{t}^{\infty} \psi(t),
 \end{equation}
 the normalization condition of Eq. (\ref{normalization}) can be expressed by
 \begin{equation}
 \Psi(0) = 1. 
 \end{equation}
 
 In the case where the waiting time distribution $\psi(t)$ used to generate the events of Fig. (\ref{sketch}) is an inverse power law with $\mu < 3$, the events separating a laminar region from the one following or from the one directly before, are called \emph{crucial events}.

 Thus the resulting aged waiting time distribution will be characterized by the following property: (a) The weight of the short times will decrease, while the weight of the long times will increase. This is essential to fit the normalization condition of Eq. (\ref{normalization}). In other words, the tail of the aged distribution density will become slower, something equivalent to decrease $\mu$. In the special case $2 < \mu < 3$, the mean value of $\tau$ of the waiting distribution density 
 \begin{equation} \label{brandnew}
 \psi(\tau) = (\mu -1) \frac{\Theta^{\mu -1}}{\left(\tau + \Theta)\right)^{\mu}}
 \end{equation}
 is given by 
 \begin{equation} \label{mean}
 \left<\tau \right> = \frac{\Theta}{(\mu -2)}.
 \end{equation}
 The waiting time distribution density of age $t_a$ is given by
 \begin{equation} \label{integral}
 \psi_{t_a}^{(ren)}(\tau) = \int_{0}^{t_a} dt' R(t') \psi(t_a + \tau - t'). 
 \end{equation}
 Due to the fact that $\left<\tau\right>$ is finite, when $t_a$ is very large we can set
 \begin{equation}
 R(t') = \frac{\mu-2}{\Theta}. 
 \end{equation}
 With simple algebra we derive from Eq. (\ref{integral})
 the following expression 
 \begin{equation} \label{integral2}
 \psi_{t_a}^{(ren)}(\tau) =  T^{\mu -2} \left\{\frac{1}{(\tau+ \Theta)^{\mu-1}} - \frac{1}{(t_a +\tau+ \Theta)^{\mu-1}}\right\}.
 \end{equation}
 Sending $t_a$ to $\infty$ we get
 \begin{equation} \label{aged}
 \psi_{t_a}^{(ren)}(\tau) = (\mu -2) \frac{\Theta^{\mu -2}}{\left(\tau + \Theta\right)^{\mu-1}},
 \end{equation}
 which can be derived from Eq. (\ref{brandnew}) by replacing $\mu$ with $\mu-1$.
 It is interesting to notice that the corresponding survival probability reads
 \begin{equation} \label{survivalaged}
 \Psi_{t_a = \infty}^{(ren)} = \left(\frac{\Theta}{\tau + \Theta}\right)^{\mu-2}. 
 \end{equation}
 The brand new survival probability 
 reads 
 \begin{equation} \label{survivalyoung}
 \Psi(\tau) = \left(\frac{\Theta}{\tau + \Theta}\right)^{\mu-1}. 
 \end{equation}

 To deepen our understanding of why the term \emph{survival probability} is appropriate to denote the function of Eq. (\ref{survivalyoung}), let us fill the times of each laminar region with either the value $+1$ or the value $-1$, according to a coin-tossing prescription.
 This creates a time series $\xi(t)$, where $\xi{t}$ has either the value of $+1$ or of $-1$. 
 Let us create a very large number of realizations $2M$, and let us select those realizations that have the first laminar region filled with 
 $+1$'s.  Then, let us make an average over all these realizations so as to create a function $\Pi(t)$ defined
 by
 \begin{equation}
 \Pi(t) = \left(\frac{1}{M}\right)\sum_{i = 1}^{M} \xi^{i}(t),
 \end{equation}
 with $M$ sufficiently large as to adopt a probabilistic approach to evaluate $\Pi(t)$. Of course, this function fits the property $\Pi(0) = 1$. By increasing $t$ the probability of finding crucial events increases. When one crucial event is met, the function $\Pi(t)$ falls to $0$, because the adoption  of the coin tossing procedure to fill the laminar region implies that half of the second laminar regions are filled with $+1$'s and half of them are filled with $-1$'s. Thus we have that
 \begin{equation}
 \Pi(\tau) = \int_{\tau}^{\infty} dt \psi(t) = \Psi(\tau). 
 \end{equation}
 
 Let us now address the issue of evaluating the stationary correlation function of this set of virtually infinitely many realizations. It is known that the adoption of renewal theory \citep{geisel} yields
 \begin{equation} \label{renewalprescription}
 \Phi(\tau) \equiv \left<\xi(t) \xi(t+\tau)\right> = \frac{1}{<\tau>} \int_{\tau}^{\infty} dt (t-\tau) \psi(t). 
 \end{equation} 
 This is so because to find the equilibrium correlation function $\Phi(\tau)$ we have to move a time window of size $\tau$ along a realization and we have to make sure that the moving window is within a laminar region, since overlapping two or more laminar regions would annihilate the correlation function. The adoption of the renewal prescription of Eq. (\ref{renewalprescription}) yields with a simple algebra and using Eq. (\ref{mean})
 \begin{equation}
 \Phi(\tau) = \left(\frac{T}{\tau + T}\right)^{\mu-2} = \Psi_{t_a = \infty}^{(ren)}(\tau). 
 \end{equation} 
 
 This observation led the authors of \citep{aging3} to study the aging of correlation function of renewal processes with the method called \emph{generalized Onsager principle}. 
 
 Although the literature on renewal aging is much more extended than this concise review, we limit ourselves to draw the reader's attention on the paper of \citep{bianco}. The authors of this paper gave a significant contribution to the assessment of the renewal condition by replacing the evaluation of $\Psi_{t_a}^{(ren)} (\tau)$ by shuffling the sequence of times $\tau$ generated by the moving window of size $t_a$. The method was successfully used to prove that systems at criticality generate renewal  events. It is important to stress that the long-time region of these crucial events is filled with Poisson events. The assessment of the renewal nature of these events would be extremely difficult, and it is to the best of our knowledge a problem not yet solved. The use the shuffling illustrated in Fig. (\ref{sketch}) allows us to assess the renewal nature or the lack of it, in spite of the fact that the analytical form of the aged survival probability is not yet known.

 \end{document}